\documentclass[11pt,epsf]{article}

\usepackage{amssymb}
\usepackage{verbatim}
\usepackage{epsfig}
\usepackage{graphicx}

\def\beq{\begin{eqnarray}}
\def\eeq{\end{eqnarray}}
\def\bea{\begin{eqnarray*}}
\def\eea{\end{eqnarray*}}

\def\centeron#1#2{{\setbox0=\hbox{#1}\setbox1=\hbox{#2}\ifdim
\wd1>\wd0\kern.5\wd1\kern-.5\wd0\fi
\copy0\kern-.5\wd0\kern-.5\wd1\copy1\ifdim\wd0>\wd1
\kern.5\wd0\kern-.5\wd1\fi}}
\def\ltap{\;\centeron{\raise.35ex\hbox{$<$}}{\lower.65ex\hbox{$\sim$}}\;}
\def\gtap{\;\centeron{\raise.35ex\hbox{$>$}}{\lower.65ex\hbox{$\sim$}}\;}
\def\gsim{\mathrel{\gtap}}
\def\lsim{\mathrel{\ltap}}

\def\singleandthirdspaced{\baselineskip=\normalbaselineskip\multiply
    \baselineskip by 130\divide\baselineskip by 100}

\newcommand{\Mp}{M_{\rm Pl}}
\newcommand{\m}{m_{3/2}}

\newcommand{\p}{\langle\phi^\dagger\phi\rangle}

\newcommand{\slsh}[1]{#1\hspace{-.15cm}/}

\begin{document}
\begin{titlepage}
\begin{flushright}
{\large hep-th/0507218 \\WIS/17/05-JUL-DPP\\
}
\end{flushright}

\vskip 1.2cm

\begin{center}

{\LARGE\bf Constraining Modular Inflation in the MSSM from Giant
Q-Ball Formation}

\vskip 1.4cm

{\large  Micha Berkooz$^a$, Daniel J.H. Chung$^b$ and Tomer Volansky$^a$}
\\
\vskip 0.4cm
{\it $^a$Weizmann Institute of Science, Rehovot, Israel}
\\
{\it $^b$Department of Physics, University of Wisconsin, Madison, WI 53706} \\

\vskip 4pt

\vskip 1.5cm

\begin{abstract}
  We discuss constraints on which flat directions can have large
  vacuum expectation values (VEVs) after inflation.  We show that
  only flat directions which are not charged under B-L and develop
  positive pressure due to renormalization group effects can have
  large VEVs of order $\Mp$.  For example, within the MSSM only the
  $H_uH_d$ flat direction is found to be viable. This strongly
  constrains the embedding of a broad class of inflationary models in
  the MSSM or some other supersymmetric extension of the SM.  For flat
  directions with negative pressure, the condensate fragments into
  very large Q-balls which we call Q-giants.  We discuss the
  formation, evolution and reheating of these Q-giants and show that
  they decay too late.  The analysis requires taking into account new
  phases of the flat directions, which have been overlooked in the
  formation and dynamics of the Q-balls.  These constraints may be
  ameliorated by invoking a short period of thermal inflation.  The
  latter, however, is viable in a very narrow window of parameter
  space and requires fine tuning.
  
\end{abstract}

\end{center}

\vskip 1.0 cm

\end{titlepage}
\setcounter{footnote}{0} \setcounter{page}{2}
\setcounter{section}{0} \setcounter{subsection}{0}
\setcounter{subsubsection}{0}

\singleandthirdspaced

\section{Introduction}
\label{sec:introduction}
Inflation is a successful cosmological paradigm both from an
observational point of view and theoretical point of view.  It is
observationally successful as the mounting cosmological data indicate
that we live in a FRW universe with a flat spatial section and that
density perturbations that seeded structure formation is adiabatic and
nearly scale invariant.  It is even partially falsifiable as a large
class of inflationary models called ``slow-roll'' inflationary models
may be ruled out through the consistency condition \cite{Sasaki:1995aw} if 
we are successful in measuring B-mode
polarization which contains information about primordial gravity waves.
From a theoretical point of view, inflation is appealing because it is
the only known way to honestly solve the flatness problem of
cosmology.  Furthermore, because the energy scale of its physics may
be as high as $10^{16}$ GeV, fundamental theories such as string
theory may have a good chance of making contact with its observable
physics.

Unfortunately, inflation does not have any smoking gun signatures yet
because it has not made any connection with controlled
experimental settings such as terrestrial colliders.  Furthermore,
there are no compellingly beautiful models of inflation as the
constraints from the requisite large number of e-foldings, the small
amplitude of density perturbations, and the smallness of the departure
of the spectral index from scale invariance necessitate unnaturally
tuned parameters in ad hoc constructions of inflationary models.  The
best hope for discovering whether inflation has occurred in our
universe's history seems to be in finding a natural inflaton candidate
in physics beyond the standard model (SM) discovered at future colliders.

One of the best motivated ideas for physics beyond the SM is
supersymmetry (SUSY), and in particular, signatures of the minimal
supersymmetric standard model (MSSM) are actively sought after in the
next generation of high energy physics experiments such as the LHC
(for a recent review in this direction, see for example,
\cite{Chung:2003fi}).  Furthermore, engineering required to construct
successful inflationary scenarios is often more natural in the context
of supersymmetry.  Among the plethora of scalars afforded by
supersymmetrizing the SM, flat directions are particularly enticing
candidates for inflatons.\footnote{Throughout this paper, we
collectively denote flat directions by $\phi$.}  Finally, from a
perturbative string theory point of view, one of the simpler
embeddings of the SM into the MSSM arises from gravity mediated SUSY
breaking context.  Hence, we are naturally led to ask which flat
directions in the gravity mediated SUSY breaking scenario of MSSM may
be part of the inflaton dynamics.

To this end, the story of Q-balls, a class of nontopological solitons 
\cite{Lee:1991ax,Coleman:1985ki,Lee:1988ag}, plays a crucial
role. Because nearly all of the flat directions in the MSSM have an
accidental $U(1)$ symmetry at the renormalizable operator level, they
fulfill one of the conditions for the existence of Q-balls (that of
having a conserved global charge) \cite{Kusenko:1997zq,Dvali:1997qv}.
The other crucial condition for the existence of Q-ball is that the
scalar field potential $U$ satisfies
\begin{equation}
\min_\phi \left[ \frac{U}{|\phi|^2} \right] < \frac{1}{2}
U''(0).
\label{eq:qballpotcond}
\end{equation}
This condition also has a good chance of being satisfied for MSSM flat
directions because its interactions with the heavy degrees of freedom
renormalize the mass (which is of the order of the SUSY breaking gravitino mass
$m_{3/2}$) of the flat direction as $m_{\phi}^2 \left[ 1+ K
\ln(\frac{|\phi|^2}{M^2 }) \right] |\phi|^2 $, and if $K<0$,
Eq.~(\ref{eq:qballpotcond}) can be satisfied 
\cite{Enqvist:1997si}.\footnote{Integrating out
many heavy degrees of freedom during inflation is generic since the
MSSM flat directions can be displaced far from the minimum, inducing
large masses to the fields to which it couple.}

This story finds relevance in cosmology \cite{Kusenko:1997si} because
for situations in which a flat direction is displaced far from the
origin at the end of inflation, one can show that when $K<0$, there is
negative pressure \cite{Turner:1983he}, which generates the necessary
instability for Q-ball formation.  Furthermore, since the $U(1)$ is an
accidental symmetry, non-renormalizable operators which break this
$U(1)$ induce large charge in the condensate which ends up inside the
Q-balls.

Therefore, as is known 
\cite{Enqvist:1997si,Kusenko:1997si,Enqvist:1998en,Enqvist:2000gq,Kasuya:2000wx,Enqvist:2003gh},
Q-balls generically form at the end of inflation if a flat direction
is turned on during inflation and in particular, if it plays a role in
the inflationary dynamics. It is also known that if the Q-balls
are too large and survive certain violent destructive processes, they
will live a long time before decaying, leading to a reheating
temperature too low to be compatible with cosmology.  Hence, certain
MSSM flat directions can be ruled out as  candidates for the
inflaton (or more precisely, flat directions cannot be displaced far
from the origin at the end of inflation) if the Q-balls form along
these directions and live a long time.

We show that Q-balls with VEVs of order the Planck scale (which we
call ``Q-giants'' due to their attendant large charge) generically
live a long time, thus eliminating many flat directions from the
possibility of being displaced far from the minimum at the end of
inflation.  The new physical observation that allows us to make this
conclusion is that {\em between} the Q-giants, the density of gas of
flat direction particles is necessarily large, thereby inducing large
masses to all particles such that the decay rate of the gas is highly
suppressed.  Consequently, not only do the Q-giants live longer due to
their suppressed decay rates, but also the thermal plasma produced by
the decay of the {\em gas} is too cold to dissociate the Q-giants.
Thus, all MSSM flat directions with negative value of $K$ are excluded
as inflaton candidates.  The only viable flat direction of the MSSM
that does not carry B-L charge is found to be $H_u H_d$.  Finally,
although B-L charge density carried by the effective Affleck-Dine
field with VEV close to $\Mp$ is generically too large to be
consistent with cosmology even when no Q-giants form, such directions
may survive due to charge dilution through mechanisms such as thermal
inflation (although as we discuss, constructing such thermal
inflationary models will be contrived and requires fine tuning).
Thus, $H_uL$ direction which has positive pressure may survive as an
inflaton candidate.

Although most of the paper is concerned with the MSSM flat directions
having $\Mp$ field variations (hence, from a string theory point of
view, these would naturally be moduli fields), it is important to note
that most of the results can be applied to other systems in which
there is Q-giant formation starting from a field condensate.  In
particular, the constraints hold for flat directions in any extension
of the SM, such as GUT or string embedding.  Indeed, there are
realizations of inflation where some of the moduli fields cannot be
part of the MSSM, and should therefore arise from flat directions of
some extension \cite{guthrandall}.  The results of this paper may
strongly constrain a given extension.  Furthermore, during (F-term)
inflation, flat directions generically obtain corrections which may
drive them alway from the origin \cite{Dine:1995uk,Dine:1995kz}.  The
existence of such directions, even without playing a role in
inflation, is highly constrained in the same manner.

The outline of the paper is the following. Section \ref{sec:cosm-scen}
contains an executive summary of the cosmological scenario and the
results.  Section \ref{sec:framework} contains a discussion of the
framework and some of the models of inflation to which it
applies. Sections \ref{sec:negat-press-flat} and
\ref{sec:evolution-q-balls} discuss the case of negative pressure flat
directions. Section \ref{sec:negat-press-flat} deals with the
formation of Q-giants after inflation and section
\ref{sec:evolution-q-balls} deals with their subsequent evolution leading to too low of a reheat temperature to be compatible with cosmology.
In section \ref{sec:posit-press-black} we discuss positive pressure
flat directions and show that they are cosmologically
acceptable. Section \ref{sec:discussioncandcaveats} deals with thermal
inflation as a possible loophole to our results.  We argue that such a
scenario however requires fine tuning and is very restricted.  We
conclude in section \ref{sec:conclusions}.

We will use the notation $\Mp \equiv 1/\sqrt{8 \pi G_N} = 2.4 \times
10^{18}$ GeV throughout the paper.

\section{The Cosmological Scenario: An Executive Summary}
\label{sec:cosm-scen}

In this section, we briefly outline the cosmological scenario and the main 
results of this paper.

We study inflationary models in the framework of gravity mediated
supersymmetry breaking.  In many natural models of inflation
an inflationary epoch is induced with the use of moduli whose 
VEVs
are $\Mp$ away from the true minimum and which dominate the energy
density during inflation.  A canonical example is hybrid
inflation \cite{Linde:1991km,guthrandall,Berkooz:2004yc,Binetruy:1996xj,Halyo:1996pp,Stewart:1996ey}.
Such moduli are usually related to flat directions in a moduli space
of a supersymmetric theory.  The purpose of this paper is to
investigate the dynamics of such flat directions and thus place
constraints on their possible role in inflation.  Furthermore, in many
theories flat directions are driven away from the origin due to
Hubble-induced couplings \cite{Dine:1995uk,Dine:1995kz}.  The existence
of such flat directions may again be constrained.  We concentrate on
the MSSM, but stress that the analysis holds for any supersymmetric
model with gravity mediated SUSY breaking.

\subsection{Assumptions}
\label{sec:assumptions}

Let us spell out the basic assumptions for which our scenario holds:
\begin{itemize}
\item At the end of inflation, the energy density is dominated by a
  flat direction, $\phi$, which is completely flat in the
  supersymmetric limit, $\m \rightarrow 0$.  As a consequence its
  potential is proportional to $\m^2$ and is therefore of the
  form,
  \begin{equation}
    \label{eq:5}
    U(\phi) = m_{3/2}^2\Mp^2 U\left(\frac{|\phi|^2}{\Mp^2},
    \frac{\phi^n}{\Mp^n} \right).
  \end{equation}
\item One consequence of the above potential is that $\phi$
  generically begins its evolution at $\Mp$.
\item Another consequence is an approximate $U(1)$ for $\langle
  \phi\rangle \ll \Mp$.  Thus at the onset of its evolution $\phi$ is
  ``side-kicked'' such that it carries an order one charge which is
  later released to the light particles including those of the MSSM.
  An immediate conclusion is that $\phi$ cannot carry B-L charge 
\cite{Berkooz:2004yc} with reheating temperature above the sphaleron
  transition temperature except under certain exceptional
  situations.\footnote{More on this point and its caveats in section
  \ref{sec:discussioncandcaveats}.}
\end{itemize}

\subsection{Cosmological Scenario}
\label{sec:cosm-high-lights}

Let us now point out the main features of the cosmological evolution
that follows from the above assumptions.  After taking into account renormalization group (RG)
corrections to the modulus potential, the latter takes the form 
\cite{Enqvist:1997si,Enqvist:1998en},
\begin{equation}
  \label{eq:84}
  U(\phi) = m_\phi^2\left[1+ K
    \log\left(\frac{|\phi|^2}{M^2}\right)\right]|\phi|^2 + \sum_{k=1}^\infty
  \frac{m_{3/2}^2 \phi^{n_k}}{\Mp^{n_k-2}} + \textrm{c.c.} 
\end{equation}
where $m_\phi\simeq \m$, $M$ is the scale at which $\m$ is defined,
typically the GUT or Planck scale and $K$ is related to the
renormalization group beta function. $K$ can be negative or positive and
typically ranges between $10^{-1}$ to $10^{-2}$.  The values of $n_k$
depend on the symmetries (possibly discrete R-symmetries) of the
model.  Initially the condensate oscillates and due to the RG
corrections positive or negative pressure is built up, depending on
the sign of $K$ \cite{Turner:1983he}.  
Before we present our computation, we summarize the $\phi$ field
evolution in each of the two cases as follows:
  
{\bf Negative Pressure, $K<0$}
\begin{itemize}
\item Negative pressure fragments the coherent condensate into very
  large Q-balls (Q-giants) and non-coherent, non-thermal particles
  with typical momentum of order $m_\phi^{-1}$.
\item While almost all of the initial charge is stored in Q-giants,
  both the Q-giants and the $\phi$ gas initially store comparable
  amount of energy.
\item Q-giant collisions break the giants into either smaller Q-balls
  or into the non-coherent gas.  We show that while the gas comes to
  dominate the energy density rather quickly, at least a percent is
  left in the Q-giants.
\item Outside the Q-giants $\langle\phi\rangle = 0$, but
  \begin{equation}
    \label{eq:28}
    \p \simeq 10^{-2}|K|^2\Mp^2.
  \end{equation}
  This finite density effect induces large masses
  to any fields which are directly coupled to $\phi$. As $\phi$ decays
  through theses heavy particles, its decay rate is suppressed by
  $1/\p$.
\item The gas reheats the light particles outside the Q-giants to
  intermediate temperature,
  \begin{equation}
    \label{eq:39}
    T_{\rm inter} \simeq 10^5 \textrm{ GeV}.
  \end{equation}
  This temperature is too low to dissociate the Q-giants, and the
  smaller than naively expected temperature is due to the decay rate
  being suppressed by $1/\p$.  
\item At this stage the temperature outside (for example, of the MSSM)
  dominates the energy density.  However, since the Q-giants still
  hold at least  a percent of the energy density in the form of
  non-relativistic matter, there is sufficient time for them to
  overcome the thermal radiation and dominate the energy density a
  long time before they decay, reheating to,
  \begin{equation}
    \label{eq:40}
    T_{\rm reheat} \simeq 10^{-4} \textrm{ GeV}.
  \end{equation}
\end{itemize}

\vspace{24pt}

{\bf Positive Pressure}
\begin{itemize}
\item In such cases, no fragmentation occurs.
\item Due to the positive pressure, a finite Jeans scale develops,
  preventing the growth of small perturbations.
\item By the time the evolution of the perturbations become
  non-linear, the Jeans scale is large enough to prevent the formation
  of primordial black holes (BHs).
  Therefore, there is no significant cosmological constraint coming
  from overproduction of BHs.
\end{itemize}

\subsection{Conclusions}
\label{sec:results}

The above dynamics strongly constrains the possible use of flat
directions in inflation:
\begin{itemize}
\item {\bf Negative pressure flat directions cannot develop VEVs of
  order $\Mp$ for viable cosmology}. They are
  excluded since they decay very late, dominating the energy density
  during nucleosynthesis thus destroying its predictions.
  Furthermore, if charged under B, they release too much baryon
  number which cannot be erased at such low temperature.
\item Positive pressure flat directions do not have the above
  problems, and further solve the problem of extensive formation of
  primordial BHs.
\item Without invoking thermal inflation
  \cite{Lyth:1995ka,Stewart:1996ai,Asaka:1999xd}, B-L directions that
  either carry B or which reheat to temperatures high enough to
  produce B from sphaleron transitions are excluded due to excess
  baryon production.  In section
  \ref{sec:discussioncandcaveats} we stress the difficulties in
  writing a model which incorporates a period of thermal inflation.
\item {\bf For the case of flat directions in the MSSM, the only
  positive pressure flat direction which is not charged under B-L is
  $H_uH_d$}.  Hence within the MSSM and excluding exceptional
  situations presented in section \ref{sec:discussioncandcaveats}, this is the only possible
  direction which can play a role in hybrid inflation or any other
  modular inflation.
\item More generally, flat directions obtain Hubble-induced
  contributions to their masses during inflation.  Such contributions drive them
  towards or away from the origin depending on the sign of the
  mass.  Hence $H_uH_d$ is the only exactly flat direction in
  the MSSM which can obtain negative contributions of this kind. 
\end{itemize}

\section{The Framework}
\label{sec:framework}

One way of classifying models of inflation is by considering whether
or not the inflaton field varies a field distance of order $\Mp$
during inflation.  This classification has phenomenological utility as
it can tell us about whether significant gravitational waves are
generated or not \cite{Lyth:1996im}.  Theoretically, large field
variations signal theories with symmetries as potentials must be
extremely flat to generically allow such large field variations
without costing excessive energy.  For example, in the context of
perturbative string theory, Planckian field variations are natural, as
the most ubiquitous scale in this theory is the string scale, and
supersymmetry and modular invariance protect flat potential
directions.  As we will see, $\Mp$ field variations are also natural
in hybrid inflationary models embedded in supergravity.

The results of our present paper regarding
the formation and longevity of Q-giants are likely to apply to most
scenarios of inflation in which the inflaton field at the end of
inflation must still travel Planckian field values to reach the
current minimum, and if the potential for this field satisfies the
condition for Q-ball formation.  The reasons is of course because the
defining characteristic of a Q-giant, large Q-charge, is due to the
large field displacement from the potential minimum.  Therefore, even
though we devote most of our attention in this paper with the end of
inflation in the context of MSSM flat directions having Planckian field
displacements, in the next subsection, we discuss more generally
possible inflationary models where our results might be applicable.
Note that although non-renormalizable operators are generically
important for the inflationary scenarios with (near-)Planck field
variation, the Q-giant analysis is insensitive to the details of the
non-renormalizable operators.

\subsection{Applicable models}
\label{sec:applicable-models}

There are many models of inflation that have been proposed which can
involve Planck scale field variations.  Arguably, most popular models
of such type are chaotic inflationary scenarios \cite{Linde:1983gd} and
hybrid inflationary scenarios \cite{Linde:1991km,Copeland:1994vg}.
Typically, for chaotic inflationary scenarios, the field variation is
of order $\Mp$ during inflation and after inflation. On the other
hand, hybrid inflation can involve field variations that are very
small during inflation while the field variation is large after the
end of inflation.

In the context of SUSY, the generic structure of SUSY
inflation comes from the scalar potential
\begin{equation}
  \label{eq:23}
  U=\frac{1}{2}\Re[f_{ab}^{-1}]D_{a}D_{b}+e^{
K/\Mp^{2}}g^{ij*}(D_{i}W)(D_{j}W)^{*}-\frac{3}{\Mp^{2}}e^{
K/\Mp^{2}}W^{*}W
\end{equation}
where
\begin{equation}
g_{ij}\equiv\partial_{\phi_{i}}\partial_{\phi_{j}^{*}}K\end{equation}
\begin{equation}
D_{i}W=\frac{\partial
W}{\partial\phi_{i}}+\frac{1}{\Mp^{2}}\frac{\partial
K}{\partial\phi_{i}}W\end{equation} and where the D-terms in the flat space limit
look like
\begin{equation}
  \label{eq:24}
  D_{a}\sim g\xi_{a}-g(\phi_{i}^{*}T_{ij}^{a}\phi_{j})
\end{equation}
with $f_{ab}$ being the gauge kinetic function and $\xi_{a}$ not
necessarily vanishing for a $U(1)$. For $U\neq0$, if $D_{a}\neq0$, we
have D-term inflation \cite{Stewart:1994ts,Binetruy:1996xj,Halyo:1996pp}, 
while otherwise,
we have F-term inflation for which $W\neq0$ or $D_{i}W\neq0$.

F-term models typically suffer from the $\eta\equiv U'' \Mp^2/U \geq
1$ problem (the $\eta$ problem) \cite{Copeland:1994vg} which arises
from the $e^{ K/\Mp^{2}}$ factor.  Although the D-term models
typically do not suffer from this problem (since the F-term is zero by
construction), in the context of string theory, if $g\xi_{a}$ comes
from an anomalous $U(1)$, its value is generically too large to be
compatible with cosmology. Note that even in D-term models the $\eta$
problem can arise since \begin{equation}
\frac{1}{2}\Re[f_{ab}^{-1}]D_{a}D_{b}\sim\xi^{2}g^{2}(1+\frac{\sigma^{2}}{\Lambda^{2}}+...)\end{equation}
which contributes\begin{equation}
\eta\sim\frac{\xi^{2}\Mp^{2}}{V\Lambda^{2}}\frac{g^{2}}{8\pi}\end{equation} 
which
would generically be too large for inflation to take place 
\cite{Lyth:1997ai}.  However, this problem can be cured by imposing
appropriate discrete symmetries.

D-term inflation typically realizes hybrid inflationary scenarios
while F-term inflationary scenarios can realize both hybrid and
chaotic scenarios.  Because chaotic inflationary scenarios generically
require the mass of the inflaton to be of order $10^{11}$ GeV, these
directions are generically heavier than MSSM flat directions
(i.e. this scenario is largely irrelevant for embedding MSSM flat
directions into an inflationary scenario). Furthermore, although this
scenario naturally will have Q-ball production, it gives rise to
sufficiently large reheating temperature for a viable cosmology.
Hence, we will now focus on discussing the salient features of hybrid
inflationary scenarios.

A common hybrid inflationary potential parameterization is 
\begin{equation}
U(\sigma,\phi)=\kappa^{2}\left(M^{2}-\frac{\phi^{2}}{4}\right)^{2}+\frac{\lambda^{2}\phi^{2}\sigma^{2}}{4}+\frac{m^{2}\sigma^{2}}{2}+\frac{\gamma}{4!}\sigma^{4}\label{eq:stdhybrid}\end{equation}
where $\sigma$ is the field direction that is rolling during inflation
when $\phi=0$ and $\phi$ is the waterfall field direction that shuts
off inflation at the end when $\sigma$ rolls down to 
$
\sigma^2
<\sigma_f^2 \equiv 2M^2 \kappa^2/\lambda^2
$.  In the standard
scenario, we have
\begin{equation}
 \kappa^{2}M^{4}\gg
m^{2}\sigma^{2}\gg\gamma\sigma^{4} \label{eq:dominance}
\end{equation} which implies that the constant
energy term $\kappa^2 M^4$ dominates the energy density during
inflation.  The dominance of the constant energy density is the
defining characteristic of hybrid inflationary models.  When combined
with the slow roll conditions and number of e-fold constraint with $m=
m_{3/2}\sim 1 $TeV and $M \equiv \beta \Mp$ (the latter says that
field variations of the waterfall field $\phi$ are Planckian at the end
of inflation), the phenomenologically allowed parameter ranges are
\begin{eqnarray}
\frac{1}{\lambda}&
<\frac{1}{\sqrt{24\pi^{2}}}\frac{\kappa^{2}\beta^{5}}{\sqrt{P_{\mathcal{R}}}}(\frac{\Mp}{m_{3/2}})^{2}\ll
& \beta\frac{\Mp}{m_{3/2}} \label{eq:cond1}\\
\sqrt{\gamma}\kappa^{3}\beta^{6} & \leq & 10^{-49} \label{eq:cond2}
\\
\frac{m_{3/2}^2}{\Mp^2} & \ll \kappa^{2}\beta^{4} \ll &
\frac{3}{8}(8 \pi)^2 P_{\mathcal{R}}~~\sim10^{-8}  \label{eq:cond3}\\
\kappa^{2}\beta^{4} &\sim
&\frac{m_{3/2}^2}{\Mp^2}\frac{N}{\ln(\frac{\sigma_{N}}{\sigma_{f}})} 
\label{eq:cond4}
\end{eqnarray}
where $P_{\mathcal{R}}\sim 10^{-9}$ is the adiabatic scalar perturbation power
spectrum, $N\sim 50$ is the number of e-folds required for inflation,
and \begin{equation}
\sigma_N \approx \frac{1}{2 \pi} \frac{1}{\sqrt{3 P_{\mathcal{R}}}} 
\frac{(\kappa^2 M^4)^{3/2}}{\Mp^3 m^2}
\label{eq:fieldatnefolds}
\end{equation} is the field value $N$ e-folds before the end of
inflation.\footnote{This is under the approximation that inflation
really ends when $\sigma=\sigma_f$ which is only approximately true
when the $\phi$ field is not too light.}  The conditions given by
Eqs.~(\ref{eq:cond1}) and (\ref{eq:cond2}) can be associated with
Eq.~(\ref{eq:dominance}), while Eq.~(\ref{eq:cond3}) comes from the
standard slow roll conditions $\epsilon \equiv \frac{ \Mp^2}{2}
(U'/U)^2 \ll 1$ and $\eta \equiv \Mp^2
(U''/U) \ll 1$.

A few remarks are now in order. 
\begin{enumerate}
\item In the absence of symmetries, the parameters $\kappa,\lambda$,
  and $\gamma$ must be finely tuned.
\item Within the context of supergravity, the waterfall field, $\phi$,
  is $\Mp$ away from its minimum at the end of
  inflation \cite{guthrandall}.  This has two consequences: (i) $\phi$
  is weakly coupled to the MSSM and therefore suffers from a moduli
  problem \cite{Berkooz:2004yc}.  One solution to this problem is to
  let the true minimum (as well as the origin of field space) be an
  enhanced symmetry point in which case the decay width of $\phi$ is
  large.  (ii) By the time $\phi$ decays, $\sigma$ has mass of order
  the Planck scale.
\item If the parameterization of Eq.~(\ref{eq:stdhybrid}) is used, the
  waterfall field $\phi$ rolls away from the origin at the end of
  inflation.  As noted by \cite{guthrandall}, this is one of the main
  reasons why it is difficult to identify one of the MSSM flat
  directions with $\phi$: MSSM flat directions have VEVs that are zero
  or much smaller than $\Mp$ today.  One may naively think it is
  easy to shift the origin of the inflationary potential.  However,
  this is incorrect since a vacuum shift generically breaks the
  underlying symmetries governing the form of the original potential.
  Nevertheless, it could well be, if one is to solve the moduli
  problem, that the light MSSM fields do live away from the origin on
  another enhanced symmetry point.  While common in string theory,
  such theories are complicated to write down, since the symmetries
  cannot be realized linearly.  One should therefore think about the
  above Lagrangian as an expansion of the full theory around the origin,
  and can thus view $\phi$ as an MSSM field rolling away from the
  origin, and into the MSSM enhanced symmetry point.
\end{enumerate}

As a concrete example in which $\phi$ is rolling away from the origin,
let us first consider one of the the models of \cite{guthrandall} (for
improvements on this type of model and further discussion see 
\cite{Berkooz:2004yc}).  They use the superpotential
\begin{equation}
W= g \frac{\sigma^2 \phi^2}{2\Mp}
\end{equation}
(where $g$ is a coupling constant) and an instanton induced soft
breaking terms with scales which are natural for gravity mediation SUSY breaking
scenario.  The resulting potential (neglecting the SUGRA corrections)
can be written approximately as
\begin{equation}
U(\phi,\sigma)\approx \frac{3}{4}\frac{\alpha^2m_{3/2}^2}{ \Mp^2}
\left(\frac{|\phi|^2}{4} -  \beta^2 \Mp^2\right)^2 + m_{3/2}^2 |\sigma|^2 +
\frac{g^2}{\Mp^2} \left(|\phi|^4 |\sigma|^2 + |\sigma|^4 |\phi|^2 \right)
\end{equation}
which can be identified with Eq.~(\ref{eq:stdhybrid}) through $\kappa
\sim \alpha m_{3/2}/( \Mp)$, $\gamma \sim g^2 |\phi|^2/\Mp^2$, and
$\lambda\sim g^2 |\sigma_N|^2/\Mp^2$.  Note that this choice of
$\kappa$ with $\alpha \sim 1/\beta$ is natural for a flat direction
field $\phi$ in supersymmetry. With this parameterization, because
$\sigma$ is rolling towards smaller values during inflation, the
condition $\sigma_N \sim 10^4 \beta^3 m_{3/2} \gg \sigma_f \sim m_{3/2}
/\lambda $ must be satisfied.  Hence, $\beta$ cannot be too
small to inflate sufficient number of e-folds, which means, as we
advertised, that Planck scale field variation of $\phi$ is natural in
the context of SUSY hybrid inflation \cite{Berkooz:2004yc}.

For an example of a model in which the waterfall field ends up at $\phi=0$,
consider the model of \cite{Stewart:1994ts}.  This model has the merit
in that it also solves the $\eta$ problem.  The supersymmetric part of
the model is defined by the superpotential
\begin{equation}
  \label{eq:26}
  W=\lambda_{1}\sigma\chi\psi_{1}+\lambda_{2}\frac{\phi^{n-2}}{\Mp^{n-2}}\phi^{2}\psi_{2}
\end{equation}
and a D-term
\begin{equation}
  \label{eq:93}
  D=\Lambda^{2}-|\chi|^{2}-|\phi|^{2}+|\psi_{1}|^{2}+n|\psi_{2}|^{2}.
\end{equation}
If $|\chi|^{2}+|\phi|^{2}\leq\Lambda^{2}$, the potential is minimized
for $\psi_{1}=\psi_{2}=0$.  Hence, ignoring for the moment the soft terms,
 the inflationary scalar
potential becomes
\begin{equation}
  \label{eq:94}
  U=\lambda_{1}^{2}|\sigma|^{2}|\chi|^{2}+\lambda_{2}^{2}\frac{|\phi|^{2n-4}}{\Mp^{2n-4}}|\phi|^{4}+\frac{g^{2}}{2}(\Lambda^{2}-|\chi|^{2}-|\phi|^{2})^{2}
\end{equation}
which inflates while $\sigma$ is large enough
\begin{equation}
|\sigma|\geq\sigma_{f}'\equiv\frac{\sqrt{n}\lambda_{2}}{\lambda_{1}}\left(\frac{\Lambda}{\Mp}\right)^{n-2}\Lambda 
\label{eq:finalfieldagain}
\end{equation}
with $\chi=0$ and the waterfall field pinned at\begin{equation}
|\phi|^{2}\approx\Lambda^{2}-n\frac{\lambda_{2}^{2}}{g^{2}}\left(\frac{\Lambda}{\Mp}\right)^{2n-2}\Mp^{2}.\end{equation}
Here we have expanded to leading order in $\lambda_{2}$.

During inflation, and taking into account soft breaking terms, the
effective potential is of the form
\begin{equation}
U=\lambda_{2}^{2}\frac{\Lambda^{2n-4}}{\Mp^{2n-4}}\Lambda^{4}+m_{3/2}^{2}\sigma^{2}.
\end{equation}
In the parameterization of Eq.~(\ref{eq:stdhybrid}), this model can be
identified with $M\sim\Lambda$ and $\kappa\sim\lambda_{2}
(\Lambda/\Mp)^{n-2} $.  Hence, to satisfy CMB fluctuation amplitude,
one find using Eqs.~(\ref{eq:fieldatnefolds}) and
(\ref{eq:finalfieldagain}), $\Lambda\sim\Mp$ and $\lambda_2 \simeq
\m/\Mp$, giving $U\sim m_{3/2}^{2}\Mp^{2}$ during inflation.


After the end of inflation, we have $\chi$ rolling out to $\Lambda$
while $\phi$ rolls to 0. Note however, that due to the D-term,
$\phi$ is not a flat direction of the full gauge group.  Thus, if one
is to embed the MSSM in this model, one must introduce an additional
$U(1)$ gauge group.  As a final remark, we point out that the solution
of \cite{Stewart:1994ts} to the $\eta$-problem involves having $W =
W_\phi = 0$ which means that $\phi$ is F-flat.  Thus these class of
models may  incorporate flat directions of the MSSM or any other
supersymmetric extension.

In all of these models, there exists a global $U(1)$ which is important
for the formation of Q-balls as we later explain.  We will see in
section~\ref{sec:flat-directions-mssm}, these accidental $U(1)$
symmetries are generic in MSSM flat directions.

\subsection{A Specific Example - Flat Directions in the MSSM}
\label{sec:flat-directions-mssm}

Flat directions are ubiquitous in supersymmetric theories.  In
particular there are numerous such directions in the MSSM, as was
systematically found in \cite{Gherghetta:1995dv,Dine:1995kz}.  As
discussed in the introduction, our goal is to constrain the embedding
of moduli in models of inflation in the MSSM and more generally the
existence of flat directions in models which involve an inflationary
epoch. One such restriction was explained in \cite{Berkooz:2004yc}.
Indeed, if we are in a situation where all the energy in the universe
after inflation is stored in some field with large VEV, then this
field cannot carry B-L number, or else too much baryon asymmetry is
generated.  This kind of reasoning assumes, of course, no fine
tuning,\footnote{Finely tuned initial conditions could easily allow for
an eccentric orbit which would suppress the produced baryon number.}
although the amount of this fine tuning is not clear once fluctuations
are taken into account, due to processes similar to tachyonic
preheating.  Thus we are left with only five possible flat directions
for the waterfall field:
\begin{center}
\begin{tabular}{l|r}
  & $B+L$\\
  \hline \hline \\
  $H_uH_d$ & 0\\ \\
  $Q\bar u L\bar e$ & 0\\ \\
  $Q\bar u Q\bar d$ & 0\\ \\
  $QQQL$ & +2\\ \\
  $\bar u\bar u\bar d \bar e$ & -2
\end{tabular}
\end{center}

Further constraints on the embedding may be pursued by investigating
the cosmological evolution of such flat directions.  Indeed it is
known that under certain conditions (i.e. negative pressure and such
flat directions being charged under some global $U(1)$), they
fragment to form Q-balls.  To see that such a $U(1)$ exists for the
above directions, we note the following.  In the MSSM, the above
directions are flat at the level of the renormalizable superpotential,
and one expects these flat directions to be lifted by
non-renormalizable terms.  However, in the models discussed in
section~\ref{sec:applicable-models}, in order for the flat direction
to be naturally  $\Mp$ away from its minimum, it must be an exact
flat direction in the supersymmetric limit \cite{Berkooz:2004yc}.
This in particular means that soft breaking terms are proportional to
$\m^2$ \cite{Dine:1995kz}.  Hence assuming for simplicity only one
flat direction, denoted by $\phi$, its potential is therefore,
\begin{equation}
  \label{eq:57}
  U(\phi) = m_{3/2}^2\Mp^2\, U(\phi^n/\Mp^n,|\phi|^2).
\end{equation}
Thus as long as $\phi$ does not couple to light fields,\footnote{Light
fields means that these fields are not integrated
out.} it has an approximate $U(1)$ for $\phi/\Mp <
1$.  For the directions not charged under B or L, this is no longer
true once other fields become light.  However, as we will see in
section \ref{sec:stability-gas-phi}, with the exception of $H_uH_d$,
this only occurs very late in the cosmological evolution, due to $\p$
condensate. For the $H_uH_d$ direction the above does not hold, due to
the $\mu$-term which gives a mass to $\phi$ already in the
supersymmetric limit.  Here we do not attempt to solve the
$\mu$-problem and assume either $\mu$ is of order $\m$ or it is
promoted to a field as in the NMSSM \cite{Nilles:1982dy}.  Due to the
special properties of this direction (positive pressure), the absence
of an approximate $U(1)$ is irrelevant and as we show below it is most
suitable to play the roll of the modulus in inflation.

A remark is now in order.  Above we have discussed the approximate
$U(1)$ in the scalar sector.  One might worry that outside Q-balls,
$\phi$ may decay into fermions which remain light.  However, as we
show below by appealing to supersymmetry, since bosons which interact
directly with $\phi$ become heavy, the fermions also becomes heavy and
therefore the Q-balls remain stable.  Hence below we assume that along
the flat direction the interaction to lighter SM fields (to which it
can decay) is through some particles whose mass is proportional to the
VEV of $\phi$ or $\p$.

Flat directions are constrained even without driving inflation.
Within the context of supergravity, flat directions obtain
Hubble-induced corrections to their masses,
\begin{equation}
  \label{eq:92}
  U(\phi) = -cH_I^2|\phi|^2 + ...
\end{equation}
during inflation \cite{Dine:1995uk,Dine:1995kz}. Here $H_I$ is the
Hubble constant at inflation and $c$ is a model-dependent
dimensionless parameter.  If $c>0$, $\phi$ becomes tachyonic during
inflation and is quickly driven away from zero to a minimum which, for
exact flat directions, is $\Mp$ away.  It then remains at this minimum
until the end of inflation, when it starts rolling back towards the
origin.  This situation is constrained by the analysis below, much
like the embedding of modular inflation.

\section{Negative Pressure Flat Directions - Formation of Q-Giants}
\label{sec:negat-press-flat}

This section is devoted to analyzing the matter content right after
the condensate fragments. As we show, the universe contains both
Q-giants and a non-thermal distribution of low momenta $\phi$
particle.\footnote{This is in contrast with the thermal effects
discussed in \cite{Kolb:2003ke,Yokoyama:2004pf}.} This will serve as
an initial configuration for the system which we evolve through the
reheating period in section \ref{sec:evolution-q-balls}.

In the case of gravity mediated SUSY breaking, the potential along a
flat direction is given by \cite{Enqvist:1997si,Enqvist:1998en,Enqvist:2000gq,Enqvist:2003gh}
\begin{eqnarray}
  \label{eq:18}
  U(\phi) = m_\phi^2\left[1+ K
  \log\left(\frac{|\phi|^2}{M^2}\right)\right]|\phi|^2 + ...
\end{eqnarray}
where $M$ is the scale at which $m_\phi \simeq \m$ is defined,
typically the GUT scale and '...' denote non-renormalizable terms with
$\Mp$ suppressed couplings. $K$ encodes the running of the mass of the
flat direction and is determined by the RG equations (which we discuss
at greater length in section \ref{sec:posit-press-black}), therefore
it depends only on the low energy theory below the Planck scale.

As is well known (see e.g. \cite{Turner:1983he}) and rederived in
appendix \ref{sec:viri-theor-equat}, the above
quantum corrections to the flat directions result with either positive
or negative pressure, depending on the sign of $K$,
\begin{equation}
  \label{eq:48}
  \frac{P}{\rho} = \frac{K}{2+K}.
\end{equation}
If negative, the condensate at the end of inflation fragments and
forms
Q-balls \cite{Enqvist:1997si,Enqvist:2000gq,Multamaki:1999an,Kasuya:2000wx,Enqvist:2003gh}.
For the models we are interested in, this process happens relatively
quickly after inflation, when the VEV of the modulus is of order the
Planck scale. Since the charge of Q-balls is proportional to the
square of the VEV,
these large charged Q-balls can be called Q-giants which decay late,
producing a low reheat temperature (and in some cases, too much baryon
number). Thus negative pressure flat directions result with
unacceptable cosmology and therefore cannot play the role of the
inflationary moduli.

After reviewing some known properties of Q-balls in subsection
\ref{sec:q-balls}, we show in \ref{sec:q-balls-formation} that the
condensate fragments into Q-giants (whose statistics we compute) and a
non-thermal distribution of $\phi$ particles. Section
\ref{sec:stability-gas-phi} discusses the properties of this gas.

%

\subsection{Q-Balls Properties}
\label{sec:q-balls}

Let us briefly summarize the basics of non-topological solitons,
otherwise known as Q-balls \cite{Lee:1991ax,Coleman:1985ki} (for a
review see e.g.  \cite{Enqvist:2003gh} and references therein).
Consider a complex scalar field with a $U(1)$ invariant potential $U(|\phi|)$,
\begin{equation}
  \label{eq:7}
  {\cal L} = |\partial_\mu\phi|^2-U(|\phi|).
\end{equation}
The conserved charge is given by $Q = i\int
d^3x(\dot\phi^\dagger\phi - \phi^\dagger\dot\phi)$.
We choose an ansatz of the form, 
\begin{equation}
  \label{eq:8}
  \phi = \phi(r)e^{i\omega t},
\end{equation}
under which the equation of motion (EOM), charge and mass become,
\begin{equation}
   \label{eq:10}
  \frac{d^2\phi}{dr^2}+\frac{2}{r}\frac{d\phi}{dr} - \phi\frac{dU_{\rm eff}}{d\phi^2} = 0,
\end{equation}
\begin{eqnarray}
 \label{eq:11}
  Q =  2\omega\int d^3x\phi^2(r),
\end{eqnarray}
\begin{eqnarray}
  \label{eq:12}
  M_Q = \int d^3x \left[(\partial_r\phi)^2 + U_{\rm
  eff}(\phi)\right] + \omega Q,
\end{eqnarray}
with $U_{\rm eff}(\phi) = U(\phi)-\omega^2\phi^2$.  This EOM describes
a particle moving in an inverted effective potential with friction
$\propto 1/t$.  We seek a solution which minimizes the energy and with
boundary conditions, $\phi^\prime(0) = \phi^\prime(\infty) = 0$. A
stable solution necessarily exists if $2U(\phi_0)/\phi_0^2$
has a minimum away from the origin. 
This condition holds for the potential (\ref{eq:18}) for negative
$K$, which is the case we are dealing with here (postponing positive
$K$ to section \ref{sec:posit-press-black}).

Here and below, we assume $|K|\log|\phi|^2/M^2 \ll 1$.  This allows us
to use the 1-loop approximation for the potential without having to
sum over large logs.  We will later see that given $K$ of order
$10^{-2}$ this assumption is easily justified, at least at the onset
of Q-ball formation. 

The Q-ball solution relevant in gravity mediated SUSY breaking is the
thick wall Q-ball solution  \cite{Enqvist:1998en}. The ansatz for this solution
\begin{eqnarray}
  \label{eq:19}
  \phi(r) = \phi_0e^{-\frac{r^2}{R^2}}
\end{eqnarray}
with $\phi_0 \ll \Mp$ (so that we can ignore non-renormalizable
terms).  Plugging this ansatz in the EOM (\ref{eq:10}) we get,
\begin{eqnarray}
  \label{eq:20}
  -\frac{6}{R^2} + \frac{4r^2}{R^4} =
   -(\omega^2-m_\phi^2(1+K))-\frac{2Km_\phi^2 r^2}{R^2} + Km_\phi^2\log\frac{\phi_0^2}{M^2}.
\end{eqnarray}
Comparing terms one finds
\begin{eqnarray}
  \label{eq:21}
  R &\simeq& \sqrt{\frac{2}{|K|}}\; m_\phi^{-1},
  \\
  \omega^2 &\simeq& m_\phi^2\left(1-2K+K\log\frac{\phi_0^2}{M^2}\right),
\end{eqnarray}
and using eqs. (\ref{eq:11}) and (\ref{eq:12}),
\begin{eqnarray}
  \label{eq:22}
  Q &=& \left(\frac{\pi}{2}\right)^{3/2}2\omega\phi_0^2R^3
  \\
  \label{eq:25}
  M_Q &\simeq& 3\left(\frac{\pi}{2}\right)^{3/2}\phi_0^2R +
  \left(2-\frac{7}{2}K + 2K\log\frac{\phi_0^2}{M^2}\right)\left(\frac{\pi}{2}\right)^{3/2}m_\phi^2\phi_0^2R^3
\end{eqnarray}
where we used $\omega \simeq m_\phi$ and $1/R \ll m_\phi$.  

Two remarks are now in order.  First, note that eq. (\ref{eq:21})
requires that the Hubble radius is much larger than $m_\phi^{-1}$ for
Q-balls to form.  In particular, $H^{-1}/R \gg 1$ which leads to
$\phi_0/\Mp \ll (3K/2)^{1/2}$.  This requirement is consistent with
neglecting the non-renormalizable terms.  A second remark is that
stability is guaranteed by the fact that $K<0$.  This ensures that the
effective mass of $\phi$ increases as the VEV of $\phi$ decreases and
therefore
\begin{equation}
  \label{eq:60}
  \frac{M_Q}{Q} \simeq m_\phi\left(1-\frac{3}{4}K - \frac{3}{2}K\log\frac{\phi_0^2}{M^2}\right) < \lim_{\phi_0\rightarrow 0}\sqrt{U''(\phi)}.
\label{eq:comparewithfree}
\end{equation}
In particular, at $\phi_0 = 0$, further light degrees of freedom must
be reintroduced (recall that the above potential is obtained by
integrating out heavy particles).  Therefore, $\phi_0$ in the logs
should be replaced by the low energy scale which can be taken to be
$m_\phi \simeq 10^3$ GeV.  Since $K<0$, this implies that the mass of
the free particle [given by the right hand side of
Eq.~(\ref{eq:comparewithfree})] is much larger than the mass per charge
of the Q-ball, rendering these Q-balls stable.

\subsection{Formation of Q-Giants and $\phi$ Gas}
\label{sec:q-balls-formation}


To predict the size and number of Q-giants, we will make two
assumptions which are physically reasonable and generic. The first is
that, assuming no fine tuning, the orbit of the flat direction
is homogeneous on a Hubble patch with non-negligible $U(1)$ charge 
\cite{Berkooz:2004yc}.  The second assumption is that in all the
 relevant length scales, initially, $\delta\phi_i/\phi_i \simeq H/\Mp
 \simeq 10^{-15}$ when $\phi_i$ is of order $\Mp$ and the dynamics we
 are focusing on begins.\footnote{Strictly speaking, the size of the
 fluctuation might grow somewhat by the time the field moves from a
 VEV $\geq\Mp$ to the regime of validity of this potential. This
 growth is expected to change the estimate of $H$ at the time of
 Q-ball formation by a multiplicative factor of order 1.  In addition,
 such a process can only serve to increase $H$ which will result in an
 increase of the Q-giants size which will only make our bounds more
 stringent.}

When the condensate fragments, not all particles in the condensate
make it into the Q-giants. We will refer to the remnant particles as
the $\phi$ gas.

{\bf Q-Giants:}
The charge
(and therefore size) of these Q-balls is determined by the time at
which the first perturbation goes
non-linear \cite{Enqvist:1998en,Enqvist:2003gh},
\begin{equation}
  \label{eq:1}
  H_0\simeq {2|K|m_\phi\over \alpha}, \ \ \
  \alpha=-\log(\delta\phi_i/\phi_i)
\end{equation}
Given that $\phi$ dominates the energy density, we can estimate its
value when the Q-balls are formed,
\begin{equation}
  \label{eq:54}
  \phi_0 = \frac{\sqrt{3}H_0\Mp}{m_\phi} \simeq \frac{\sqrt{12}|K|\Mp}{\alpha}
\end{equation}
and using eqs. (\ref{eq:21}) and (\ref{eq:22}) one obtains,
\begin{equation}
  \label{eq:27}
  Q =
  \frac{2\pi^{3/2}}{|K|^{3/2}}\;\frac{\phi_0^2}{m_\phi^2}\simeq
  10^{28} \left(\frac{m_\phi}{10^3\textrm{ GeV}}\right)^{-2}
  \left(\frac{K}{10^{-2}}\right)^{1/2}\left(\frac{\alpha}{30}\right)^{-2}.
\end{equation}
The validity of this estimate was verified numerically in
\cite{Kasuya:2001hg,Kasuya:2000wx}.
Due to the large initial value of $\phi$, these Q-balls store an
extraordinary amount of charge, hence the name Q-giants.

We stress that the $\phi$ field that we are using cannot serve as an
Affleck-Dine (AD) field \cite{Affleck:1984fy,Dine:1995kz}. The reason
is that it dominates the energy density (since it has the largest VEV
possible in this model) and hence if it were the AD field, it would
have generated a very large baryon asymmetry.  Thus the baryon number
must be generated either by another AD field or by a variant of
EW-baryogenesis 
\cite{Carena:1996wj,Farrar:1996cp,deCarlos:1997ru,Laine:2000rm,Berkooz:2004kx,Hall:2005aq,Menon:2004wv}
if the
radiation temperature to be calculated below is above the electroweak
(EW) scale.

{\bf $\phi$-Gas:} The charge in the $\phi$ condensate is maximal if
the VEV moves in a circular orbit. In this case, all the charge
makes its way efficiently into Q-giants. In the generic situation we
are discussing, the orbit is not circular.
We show below that in this case, in addition to Q-giants, a
distribution of low momenta (non-coherent) $\phi$ particles
appears. We will refer to this as the $\phi$-gas. The existence of
such a component is also supported in numerical simulations of
condensate fragmentation \cite{Kasuya:2000wx}. The critical
features of this gas are discussed in section
\ref{sec:stability-gas-phi}.

\subsubsection{Efficiency of Q-Giants and Gas Formation}
\label{sec:efficiency-q-balls}

Our initial configuration is that of a condensate, for which the
perturbation on some scale grow to become non-linear. Once the
non-linear regime is formed, Q-giants form very rapidly, as does the
$\phi$ gas.  Immediately after fragmentation the universe contains a
mixture of the following components:
\begin{enumerate}
\item Energy in Q-giants - an order one fraction of the charge initially goes into Q-giants.
\item A gas of $\phi$ particles outside the Q-giants - the remaining
  energy goes initially into this gas. 
\item Light MSSM particles which result from the decay of the $\phi$
  particles - initially there are none of those.
\end{enumerate}

The physics of the fragmentation process is the following. The length
scale that becomes non-linear first is of
order \cite{Kasuya:2000wx}
\begin{equation}
  \label{eq:56}
  \frac{k}{a} \sim R^{-1} \simeq \left(\frac{|K|}{2}\right)^{1/2}m_\phi,
\end{equation}
where $a$ is the expansion factor.  Once this mode enters the
non-linear regime Q-giants form quickly, roughly one per the length scale
(\ref{eq:56}).  $\phi$ quanta that do not make it into Q-giants appear
into a non-coherent gas of $\phi$ (and anti-$\phi$) particles. Since
the entire process occurs for low momenta - of order $m_{3/2}$ up to
some powers of $K$ - we expect that the $\phi$ gas will contain only
low momenta. This picture is also verified numerically in
 \cite{Kasuya:2000wx}. Furthermore, the same simulations suggest that
most of the charge finds its way into the Q-giants rather than into the
$\phi$ gas.\footnote{An intuitive argument may be the following. The
  VEV of the field inside the Q-giant moves in a circular orbit in
  field space. This means that regions in spacetime in which the
  charge-density/energy-density ratio is maximal have an easier time to fall
  into Q-giants. The $\phi$ gas originate from regions of space in
  which the charge-density/energy-density ratio is low - such regions
  tend to be less coherent - but this also implies that the gas is
  characterized by a low charge-density/energy-density ratio.}

It is also easy to see that the amount of anti-Q-giants is very small.
Indeed, recall that the condensate is rotating in one direction with a
charge close to the maximal allowed.  To get a Q-giant with the
opposite charge, the condensate in a large piece of spacetime has to
{\it coherently} start rotating the other way, which is suppressed. It
is also unlikely that a Q-giant will form out of a gas of particles
after the condensate fragmented.  This is from phase space arguments:
a non-coherent collection of $\phi$ quanta will not be able to
assemble itself into a coherent Q-giant (or anti-Q-giant).

The total energy density in the $\phi$ condensate at the time of fragmentation
is of order $\rho_{\rm total}=m_{\phi}^2\phi_0^2$. The energy per charge
in the condensate is given by $\rho_{\rm total}/Q\simeq m_\phi/(1-\epsilon)$
where $\epsilon$ denotes the eccentricity of the orbit of $\phi$ and
is therefore less than unity.  On the other hand once Q-giants have
formed, they fulfill $M_Q/Q \simeq m_\phi$ [see eqs. (\ref{eq:22}),
(\ref{eq:25})].  Thus since most of the charge ends up inside the
Q-giants, the energy and charge density of the system's constituents after
the Q-giant formation is
\begin{eqnarray}
  \label{eq:90}
  \begin{array}{lllll}
  Q_{\rm gas}/V \simeq 0 &&&& \rho_{\rm gas} \simeq \epsilon m_\phi^2\phi_0^2
  \\ \\
  Q/V \simeq (1-\epsilon) m_\phi\phi_0^2 &&&& \rho_{Q} \simeq (1-\epsilon) m_\phi^2\phi_0^2
  \end{array}
\end{eqnarray}

Finally, for consistency of this picture we need to show that the
mode-mode interactions in the $\phi$ gas are small enough such that
the gas does not thermalize but rather remains at low momenta until it
decays.  Indeed, there are only two kinds of interactions which are of
importance to us: interactions arising from the log term in
(\ref{eq:18}) and non-renormalizable interactions coming from
corrections to the Kahler potential.  For the latter, as discussed in
section \ref{sec:flat-directions-mssm}, a n-point interaction is
suppressed by $\m^2/\Mp^{n-2}$.  Thus the rate for momentum transfer
can be estimated to be 
\begin{equation}
  \label{eq:59}
  \Gamma_{\rm MM}^{\rm Non-Ren} \simeq  n_\phi \langle \sigma_\phi v\rangle \lsim \frac{|K|^2\m^3}{\Mp^2},
\end{equation}
where $\sigma_\phi$ is the cross-section for these interactions and
the $|K|^2$ factor comes from the suppression in the number density of
the gas.  Here we have estimated $\Gamma_{\rm MM}$ for $n=4$.  Higher $n$ 
would result with lower rates.

The log interactions are more subtle.  As we discuss at length in
section \ref{sec:stability-gas-phi}, while $\langle\phi\rangle$
vanishes outside the Q-giants, the states which interact with $\phi$
remain massive due to non-vanishing $\p$.  Thus the log potential is
still valid outside the Q-giants and result with interactions
suppressed by $\p$.  Thus one finds
\begin{equation}
  \label{eq:61}
  \Gamma_{\rm MM}^{\rm Log} \lsim \frac{|K|^2\m^3}{\p}
\end{equation}
which is faster than the scattering rate eq. (\ref{eq:59}) and is of
order the decay rate for $\phi$ (see section
\ref{sec:stability-gas-phi}).  In fact, we expect these scattering to
be slower than the estimate above since 2-2 scattering (for which the
estimation was done) cannot result with an increase of the average
momenta while higher n-point scattering is further suppressed.  We
thus conclude that the gas outside the Q-giants is formed out of
equilibrium and stays that way until it decays.

Finally, light MSSM particles are absent until the gas decays. We
compute the gas decay rate in section \ref{sec:stability-gas-phi}. In
section \ref{sec:decay-q-balls} we show that the temperature of the
MSSM cannot dissociate the Q-giants.

\subsubsection{Initial Velocity and Charge Distribution in Q-Giants}
\label{sec:distribution-q-ball}

In section \ref{sec:q-ballq-ball} we briefly investigate the consequence of
Q-giant collisions.  In order to do that, we must  estimate the
distribution of velocity and charge of the Q-giants.  We do that here.

Let us denote a given Q-giant by its center $\vec q$ and radius $R$.  The momentum
current is then found to be
\begin{equation}
  \label{eq:62}
  P^i(\vec q, R) = \int_{V(\vec q,R)} d^3x
  a^3\left[\partial^i\phi^\dagger\partial^0\phi + \partial^0\phi^\dagger\partial^i\phi\right],
\end{equation}
where $V(\vec q, R)$ is the volume of the Q-giant and $x^i$ are
co-moving spatial coordinates.  To extract the distribution of momentum,
we follow the evolution of $\phi$ until fluctuations become of
order one, when $\phi = \phi_0$.  Thus expanding around the
homogeneous solution,
\begin{equation}
  \label{eq:63}
  \phi(x, t) = \phi(t) + \delta\phi(x,t)
\end{equation}
one finds to lowest order in $\delta\phi$,
\begin{equation}
  \label{eq:64}
  \langle \delta P^i(\vec q_x, R)\delta P^j(\vec q_y, R)\rangle \simeq
  |\dot\phi(t)|^2\int_{V_xV_y}d^3xd^3ya^2\frac{\partial}{\partial
    x^i}\frac{\partial}{\partial y^j}\langle\delta\phi^\dagger(x)\delta\phi(y)\rangle.
\end{equation}
The above can be estimated at the time of Q-giant formation by taking
$\delta\phi \simeq \phi \simeq \phi_0$ and estimating the typical
momentum to be of order $k\simeq R^{-1}$.  Hence we get
\begin{equation}
  \label{eq:65}
  \delta P \simeq \omega \phi_0^2 R^2
\end{equation}
which leads to the velocity distribution,
\begin{equation}
  \label{eq:66}
  \delta v \simeq \frac{\delta P}{M_Q} \simeq \sqrt{|K|}.
\end{equation}
The charge distribution can be found in a similar fashion.
Using the expression for Q given above, eq. (\ref{eq:8}), and expanding as 
in (\ref{eq:63}), one finds
\begin{equation}
  \label{eq:67}
  \langle\delta Q(\vec q_x, R)\delta Q(\vec q_y, R)\rangle \simeq |\phi|^2\int_{V_xV_y}d^3xd^3ya^6\frac{\partial}{\partial
    x^0}\frac{\partial}{\partial y^0}\langle\delta\phi^\dagger(x)\delta\phi(y)\rangle
\end{equation}
from which one can estimate,
\begin{equation}
  \label{eq:68}
  \delta Q \simeq \omega\phi_0^2R^3
\end{equation}
and therefore $\delta Q/Q \sim 1$.

Given the above, let us assume an initial Gaussian
distribution\footnote{This assumption is valid in the limit $\sqrt{K} \ll 1$
  in which case there are many Q-giants (of order $10/\sqrt{K}$) within a
  Hubble radius.},
\begin{equation}
  \label{eq:71}
  f(t_0, Q, v) = N\, \frac{e^{-(Q-\langle Q\rangle)^2/2\delta
  Q^2}}{(2\pi\delta Q^2)^{1/2}}\, \frac{e^{-v^2/2\delta v^2}}{(2\pi\delta v^2)^{3/2}}, 
\end{equation}
with $\langle Q\rangle$ given in (\ref{eq:22}).  Here $N$ is the normalization which can be approximated by assuming
most of the charge is stored in the Q-giants  \cite{Kasuya:2000wx},
\begin{equation}
  \label{eq:72}
  (1-\epsilon) m_\phi\phi_0^2 = \int dQ Q\int d^3v f(t_0, Q, v)
\end{equation}
where $\epsilon$ is the initial eccentricity which is of order one
(see eq. (\ref{eq:90})).  This gives,
\begin{equation}
  \label{eq:74}
  N \sim \frac{(1-\epsilon)}{R^3} \simeq (1-\epsilon) m_\phi^3 |K|^{3/2}.
\end{equation}
We will later use this distribution to estimate Q-giant destruction rates.

\subsection{Stability of the $\phi$ Gas}
\label{sec:stability-gas-phi}

In the non-relativistic gas of $\phi$ particles, the relevant energy
per particle are of order $m_{3/2}$. Hence during the stage of
Q-giant formation, one must have, $\langle(\partial\phi)^2 +
U(\phi)\rangle = \rho_\phi$ which when combined with
eqs. (\ref{eq:18}) and (\ref{eq:90})
 gives
\begin{equation}
  \label{eq:69}
  \p = f_{\rm gas} \phi_0^2.
\end{equation}
where $f_{\rm gas} \approx \epsilon$ at the onset of Q-giant formation
and is roughly 1 by the time the gas decays
due to energy loss of Q-giants through collisions (as
discussed in section \ref{sec:q-ballq-ball}). On the other hand
$\langle\phi\rangle\approx 0$ between the Q-giants.

We now argue the following:
\begin{enumerate}
\item All the states which are strongly coupled to $\phi$ are heavy -
  $m^2\sim \p$ - due to this large variance in the field and despite
  the fact that $\langle\phi\rangle=0$.
\item While giving large masses to all particles, $\phi$ itself
  remains light to all order of perturbation theory.  This is a
  consequence of supersymmetry.
\item As a result, $\phi$ is stable and decays late, reheating the
  MSSM particles to temperature $T_{\rm inter} \simeq m_\phi
  (\Mp/m_\phi)^{1/6}$.  This tempertature, as discussed in section
  \ref{sec:decay-q-balls}, cannot dissociate the 
Q-giants, which remain
  stable and dominate the energy density until they decay much later. 
\end{enumerate}

To get a sense for why the above is true, let us work with a specific
model. As an example, consider the renormalizable
interaction\footnote{Of course, if no renormalizable interactions
  exist and only $\Mp$ suppressed ones, the decay rate of the $\phi$
  gas is very low}
\begin{equation}
  \label{eq:38}
  W_0 = \lambda \phi X^2 + ...
\end{equation}
where $\lambda$ is an order one dimensionless coupling and '...'
stands for additional non-renormalizable terms. This is the type of
interaction we expect in the context of flat directions of the MSSM.
The scalar potential then
takes the form,
\begin{equation}
  \label{eq:41}
  U_{\rm scalar} = 4|\lambda|^2 |\phi^2| |X|^2 + |\lambda|^2|X|^4 +
  (\textrm{soft terms}) + ...
\end{equation}
and there are Yukawa couplings,
\begin{equation}
  \label{eq:42}
  -L_{\rm Yukawa} = 2\lambda \phi \Psi_X\Psi_X + \lambda X \Psi_\phi\Psi_X
\end{equation}
where $\Psi_{\phi,X}$ are chiral fermions.

\begin{figure}[t]
  \centering
  \includegraphics[height=2.5cm]{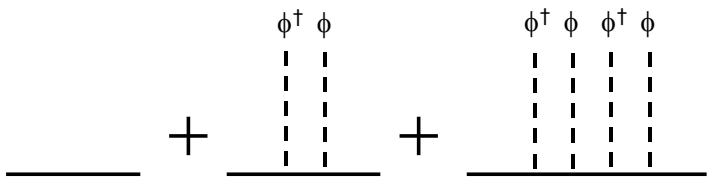}
  \caption{Diagrams contributing to the fermion's two-point
    functions.}
  \label{fig:fermion-prop}
\end{figure}

The first term in (\ref{eq:41}) gives a mass to $X$, $m_X^2 =
4|\lambda|^2\p$, thus making it very heavy.  On the other hand, one
may naively think that $\Psi_X$ remains light since it interacts with
$\phi$ rather than $\phi^\dagger\phi$.  To see that this is not the
case, recall that the physical mass for a fermion is not $m$ which is
charged under chiral rotations, but rather $|m|^2$ which in that case
will be proportional to $\p$.  Another way to see this is to look in
the limit of exact supersymmetry (assuming $\p$ does not directly
break SUSY)\footnote{It is important to distinguish 
between finite density effects (from $\p$) and finite {\em
temperature} SUSY breaking effects.}. In that limit $X$ and
$\Psi_X$ must have similar masses and therefore the fermion must be
massive as well.  To be more concrete, let us compute the new
propagator for $\Psi_X$ in the presence of non-vanishing $\p$.
Summing over the diagrams in Fig. \ref{fig:fermion-prop} one finds the
corrected propagator,
\begin{equation}
  \label{eq:43}
  \langle\Psi_X\Psi_X\rangle = \frac{\slsh{p}}{p^2- 4|\lambda|^2\p}.
\end{equation}
Indeed, the pole is at $4|\lambda|^2\p$ which is just $m_X^2$.

\begin{figure}
  \centering
  \begin{tabular}{c}
    \includegraphics[height=2.5cm]{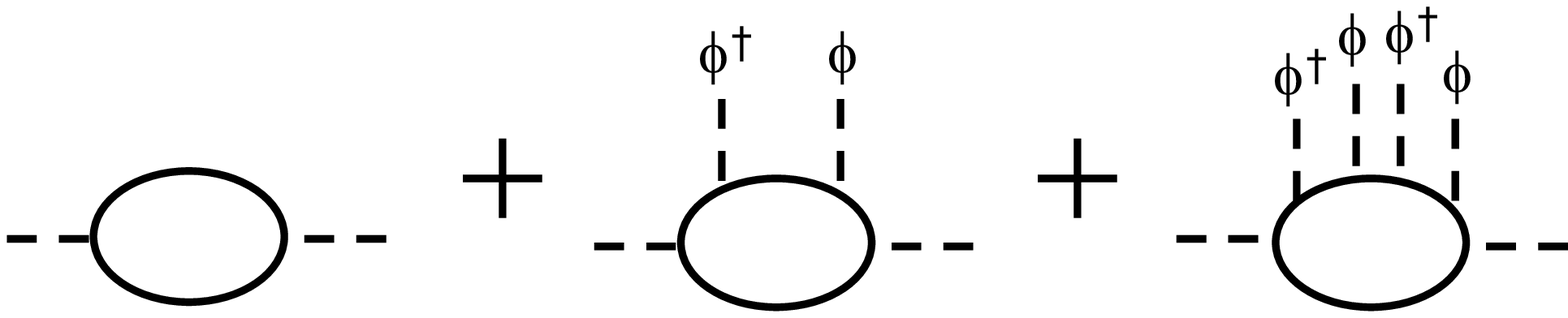} \\
    \includegraphics[height=2.5cm]{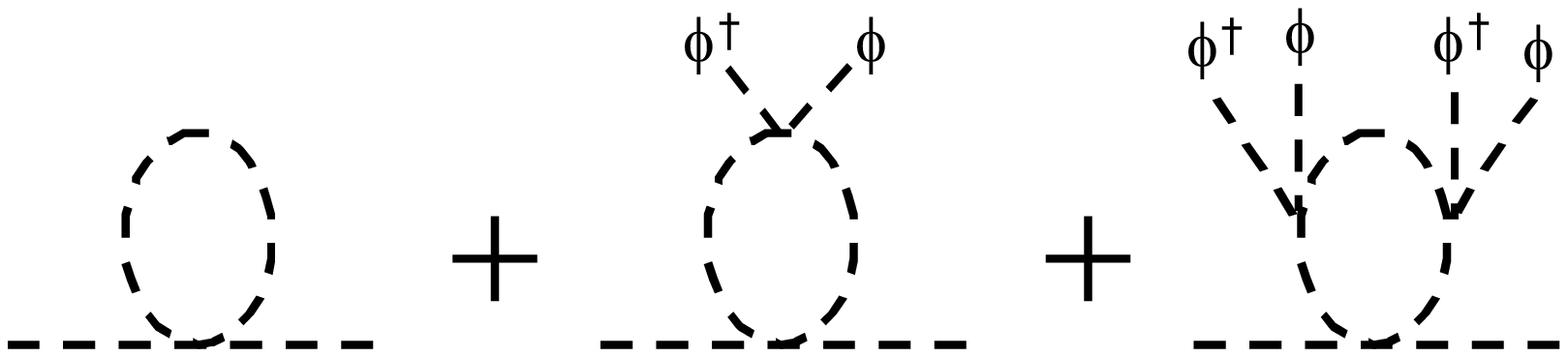}
  \end{tabular}
  \caption{Contributions of bosons and fermions to $m_\phi^2$.   In
    the supersymmetric limit, the bosonic contribution cancels the
    fermionic contribution exactly.}
  \label{fig:phi-prop}
\end{figure}

One may also worry that thermal and plasma effects induce a large
splitting between the masses of $X$ and $\Psi_X$ which contributes to
the mass of $\phi$, $m_\phi^2 \propto m_X^2 - m_{\Psi_X}^2$.  Again,
this can be argued not to occur.  Indeed, in the supersymmetric limit,
as was argued above, there is no splitting between $X$ and $\Psi_X$
and therefore no contribution the $m_\phi$.  This can also be seen, if
one notices that the contribution of the bosonic diagrams to
$m_\phi^2$ in the presence of $\p$ (Fig. \ref{fig:phi-prop}a), cancel
the contribution from the fermionic diagrams
(Fig. \ref{fig:phi-prop}b).  Any supersymmetric breaking effect, arise
either from soft breaking terms which are proportional to $m_\phi \sim
\m$ or from plasma effects (recall that the gas is not in thermal
equilibrium and therefore one cannot assign a temperature to it) which
must be of order $k \sim R^{-1} = \sqrt{|K|/2} m_\phi$.  Thus any
splitting must also be of order $m_{3/2}$ and therefore $\phi$ remains light.

Having established a consistent picture, we may now ask when does the $\phi$
gas decay.  Since it is coupled to light degrees of freedom only
through heavy particles of mass $\sim \p$, its decay rate can be
bounded by
\begin{equation}
  \label{eq:44}
  \Gamma_\phi \leq \frac{m_\phi^3}{8\pi\p}.
\end{equation}
The energy densities in the gas and in the Q-giants are comparable and
redshifts like $a^{-3}$.  Thus we may assume the gas dominates the
energy density and compute the intermediate reheating process by
solving the equation for the energy density in the radiation,
\begin{equation}
  \label{eq:47}
  \dot \rho_{\gamma} + 4H \rho_{\gamma} = \Gamma_\phi \rho_\phi.
\end{equation}
Using (\ref{eq:28}) and (\ref{eq:44}) and taking $H = 2/(3t)$ for this
matter dominated epoch, one can estimate, 
\begin{equation}
  \label{eq:70}
  \rho_{\gamma} \simeq {3\over 11}\frac{f_{\rm gas}}{8\pi} m_\phi^5 t
\end{equation}
This solution is valid until radiation becomes dominant at $H =
\Gamma_\phi$.  At this stage, the reheating ends
and the gas decays completely, producing an intermediate temperature
given by
\begin{equation}
  \label{eq:45}
  T_{\rm inter} \simeq g_*^{-1/4} m_\phi \left(\frac{f_{\rm gas}\Mp}{8\pi m_\phi}\right)^{1/6} \simeq 10^5 \textrm{ GeV}
\end{equation}
with the gas decaying when,
\begin{equation}
  \label{eq:46}
  \p \Bigg|_{\rm decay} \simeq m_\phi^2 \left(\frac{\Mp}{8\pi
  f_{\rm gas}^{1/2} m_\phi}\right)^{2/3}.
\end{equation}
Note that unlike the usual inflaton scenario, this intermediate
reheating does not start at high temperature and therefore there is no
danger to the stability of the Q-giants at this stage.

From that point onwards, radiation surrounds the Q-giants.  Initially
it dominates the energy density but very quickly (once the expansion
factor grows by two orders of magnitude) it becomes sub-dominant with
Q-giants dominating until they decay. Note that taking into account
sphalerons in the case where the charge in the giants corresponds to
B+L, is not expected to change the final B+L charge since the gas
outside the Q-giants stores very little charge.
Below we further
investigate the dynamics of the Q-giants and show that the temperature
of radiation does not dissociate the Q-giants.

\section{Evolution of Q-Giants and Reheating}
\label{sec:evolution-q-balls}

The evolution and stability of Q-balls has been widely investigated
 \cite{Cohen:1986ct,Kusenko:1997si,Enqvist:1998en,Enqvist:2003gh,Multamaki:2000qb}.
There are many effects which can influence the time of decay and the
reheat temperature.  We will consider the most relevant effects:
fragmentation of Q-giants through collisions,
dissociation due to external thermal plasma, and evaporation to bosons
and fermions.

In addition, there is potentially a new effect which is the decay of
Q-giants when immersed in the non-thermal gas of low momenta $\phi$
particle. However, as we saw above, in cases where the Q-giants are
large the gas outside has large $\p \simeq
f_{\rm gas} \phi_0^2$ which suppresses interactions between the
Q-giants and the $\phi$ gas at the surface (as well as with light particles).

\subsection{Q-Giant Collisions}
\label{sec:q-ballq-ball}

Q-giants may collide between themselves leading to charge and energy
transfer and may be the cause of Q-giants disintegration.  Collisions
are especially important since the reheat temperature is inversely
proportional to a power of the charge (see equation (\ref{eq:52})), and hence if the
Q-giants fragment to Q-balls with lower charge, the reheat temperature will be
higher.  We now show that Q-giant collisions only mildly reduce the
number of Q-giants with large charge. 

The only long range Q-giant/Q-giant interaction is gravitational. In
addition they interact through direct collisions and interact with the
gas around them. Here we focus on the effects of collisions. We expect
that the cross-section for collision is of order the size of the
Q-giant, which is verified in numerical simulations
 \cite{Multamaki:2000qb}.  More precisely, in general collisions are
divided into three types: fusion, charge exchange and elastic.  The
cross-section for each is different with charge exchange being the
largest, of order four times the geometric cross-section\footnote{This
  is the largest cross-section for a process that can reduce the
  charge. A larger cross-section can occur for fusion, but this will
  make the Q-giants larger. The cross-section for fusion strongly
  depends on the phase correlation between colliding Q-giants.  Q-giants
  with a similar phase will tend to merge, while the fusion
  cross-section for randomly distributed phases is highly
  suppressed.}.  To see how collisions reduce the number of Q-giants,
let us define
\begin{equation}
  \label{eq:73}
  n_Q(t) \equiv \int_{\langle Q\rangle /2}^{3\langle Q\rangle/2}dQ'\int d^3v f(t, Q', v)
\end{equation}
with $f(t, Q, v)$ is the distribution function in velocity $v$ and
charge $Q$ of the Q-giants (as defined in (\ref{eq:71})). We
assume that whenever there is a collision between Q-balls, they
fragment completely. With this assumption we are underestimating the
number of Q-balls, and in particular the large Q-giants (hence
obtaining an upper bound on the reheat temperature).

Within this approximation, the equation of motion
for $n_Q$ is given by,
\begin{equation}
  \label{eq:75}
  \frac{dn_Q}{dt} + 3Hn_Q = -n_Q^2\sigma v
\end{equation}
with $\sigma$ estimated to be of order the geometric cross-section
$\sigma \simeq \pi R^2$, which is approximately constant.  In the term on
the right hand side we have neglected some order one coefficient which
is related to the fact that the decay width is proportional to the
total number of Q-balls and not just the Q-giants defined in
(\ref{eq:73}).  As was
shown in section \ref{sec:distribution-q-ball}, we may take the
initial velocity, which redshifts like $a^{-1}$, to be of order $v_0
\sim \sqrt{|K|}$.  Thus using (\ref{eq:1}), one finds
\begin{equation}
  \label{eq:76}
  n_Q(t) \simeq n_Q^0 \left(\frac{a_0}{a}\right)^3\left[1 +
  \frac{2\pi\alpha}{5}\,\frac{n_Q^0}{m_\phi^3
  |K|^{3/2}}\left(1-\left(\frac{a_0}{a}\right)^{5/2}\right)\right]^{-1} \equiv n_Q^0
\left(\frac{a_0}{a}\right)^3 \epsilon_{\rm coll}(a)
\end{equation}
where $n_Q^0 \simeq \epsilon m_\phi^3 |K|^{3/2}$ is the initial number
density for Q-giants with charge of order $\langle Q\rangle$.  We see
that collisions become less effective as the universe expands.  This
is easily understood since, as the number density of the Q-giants drops
down, collisions become more rare.  Taking again $\alpha \simeq 30$,
we learn that collisions reduce the number of Q-giants by roughly two
orders of magnitude, namely
\begin{equation}
  \label{eq:77}
  \epsilon_{\rm coll}(a\rightarrow\infty) \gsim 10^{-2}.
\end{equation}

Thus even if we assume that collisions break Q-giants only into $\phi$
gas (without small Q-balls), we see that
still a few percent of the energy density is left in the Q-giants
before the $\phi$ gas decays.  Thus we may take, 
\begin{equation}
  \label{eq:78}
  f_{\rm gas} \simeq 1 -  \epsilon_{\rm coll}(a\rightarrow\infty),
\end{equation}
[see eq. (\ref{eq:69})].

Since we have assumed that the collisions cause the Q-giants to
fragment, this is a lower bound on the number of Q-giants. Two
additional mechanisms change the survival rate of the Q-giants upwards
by a small amount:

1. The field theory binding energy of the $Q$-balls is of order
   $|K|M_Q$, where $M_Q$ is the mass of the Q-ball. The kinetic energy
   is initially of order $M_Qv^2/2\simeq M_Q K$. The kinetic energy,
   however, redshifts rapidly and very quickly there is not enough
   kinetic energy to overcome the binding energy. This will increase
   the survival rate of the Q-giants by a factor of order 1.

2. In the simulations \cite{Multamaki:2000qb} Q-balls often do
not fragment upon collision. In fact, for the Q-giants the gravitational
binding energy makes them even more resistant to fragmentations since
the gravitational energy,
\begin{equation}
  \label{eq:58}
  |E_{\rm grav}| \simeq \frac{m_\phi^2Q^2}{\Mp^2 R}\simeq
  \frac{10^4}{\alpha^4}\,\frac{|K|^{3/2}\Mp^2}{m_\phi},
\end{equation}
is proportional to $Q^2$ and is similar in order to the binding energy
of the field in flat space (see (\ref{eq:90})) which is
\begin{equation}
   |E_{\rm binding}| \simeq |K|m_\phi Q \simeq
  \frac{10^2}{\alpha^2}\,\frac{|K|^{3/2}\Mp^2}{m_\phi}.
\end{equation}
$E_{grav}$ is proportional to
$Q^2$, which means that large Q-giants are energetically
favored over several smaller Q-giants of the same charge. However, for
$\alpha\simeq 30$, this effect is rather small.

The conclusion is that by the time the collisions end - shortly after
the Q-giants form - about $1\%$ of the total charge and energy is in
non-relativistic Q-giants, and the remaining charge and energy is in
the $\phi$ gas. Hence, when the gas decays, there is a short period
of radiation domination, before the Q-giants become dominant again.  As
we show in the following sections, given the intermediate
temperature, eq. (\ref{eq:45}), there is enough time between the
decay of the gas and the decay of the Q-giants for the Q-giants to
take over and dominate of the energy density.

\subsection{Dissociation of Q-Giants}
\label{sec:decay-q-balls}

After the $\phi$-gas decayed, the universe can be thought of as two
subsystems composed of Q-giants and radiation.  Because the gas never
reached thermal equilibrium before decaying, the two systems are not
in thermal equilibrium.  Nevertheless, the effects of thermal
radiation may be important and must be taken into account.  We now
compute a bound on the temperature outside the Q-giants below which,
the Q-giants dissociate.  To do so, we briefly repeat the considerations
presented in  \cite{Enqvist:1998en}.

Thermal particles can penetrate Q-giants thereby transferring energy
and possibly causing dissociation.  The condition to evade
dissociation is that the excess energy per unit charge (which was not
radiated away) is smaller than the binding energy per unit charge $\delta m \sim |K|m_\phi$:
\begin{equation}
  \label{eq:31}
  \left.\frac{\Delta M_Q}{Q}\right|_{\delta t_r} < \delta m
\end{equation}
where $\delta t_r$ is the time scale over which the Q-giants radiates
its excess energy away.  The flux of particles hitting the Q-giant is
\begin{equation}
  \label{eq:32}
  \Gamma = \frac{g_*(T)}{\pi^2} 4\pi r_{st}^2 T^3
\end{equation}
where $r_{st}$ is the stopping radius of the Q-giant defined by
$g\phi(r_{st}) \simeq 3 T$ with $g \simeq {\cal O}(1)$ being the coupling constant of
$\phi$ to radiation and $g_*$ is the number of relativistic
degrees of freedom coupled to the Q-giant.  Taking the energy per
thermal particle transferred to the Q-giant be $\gamma_TT$, we get
\begin{equation}
  \label{eq:33}
  \frac{dM_Q}{dt} = M_Q\Gamma = \frac{4 \gamma_T g_*(T)}{\pi}  r_{st}^2 T^4.
\end{equation}
$r_{st}$ can be easily calculated by using the ansatz (\ref{eq:19}),
\begin{equation}
  \label{eq:34}
  r_{st} = \beta R, \,\,\,\, \beta=\log^{1/2}\left(\frac{g\phi_0}{3T}\right).
\end{equation}
Thus, using (\ref{eq:21}) for $R$ we get, 
\begin{equation}
  \label{eq:35}
  \frac{dM_Q}{dt} = \frac{8 \gamma_T g_*(T)\beta^2}{\pi |K|m_\phi^2}T^4,
\end{equation}
which, from the inequality (\ref{eq:31}) and using $\delta t_r \simeq
1/m_\phi$ one obtains
\begin{equation}
  \label{eq:36}
  T \leq 10^8
  \left(\frac{K}{10^{-2}}\right)^{1/2}\left(\frac{200}{g_*}\right)^{1/4}\left(\frac{m_\phi}{10^3\textrm{
  GeV}}\right)\left(\frac{Q}{10^{28}}\right)^{1/4} \textrm{ GeV},
\end{equation}
where we have taken $\gamma_T\simeq 1$ and $\beta \simeq 4$.


In  \cite{Enqvist:1998en} this inequality is used as a bound on the reheat
temperature.  Here, we use it as a bound on the maximal temperature
outside the Q-giants.  It is therefore immediately clear that the
radiation after the decay of the gas cannot dissociate the Q-giants.
Below, we also investigate the reheating process during the decay of
the Q-giants themselves.  We show that dissociation does not play a
role at that stage either.

\subsection{Reheating}
\label{sec:reheating}

In general Q-balls can decay into fermions and bosons through the
surface where these particles are
light \cite{Cohen:1986ct,Enqvist:1998en}.  The decay into fermions was
found to be bounded by \cite{Cohen:1986ct}
\begin{equation}
  \label{eq:29}
  \frac{dQ}{dtdA}\leq \frac{\omega_0^3}{192\pi^2},
\end{equation}
with $\omega_0^2 = \omega^2 - m_\phi^2(1+K) \simeq 3m_\phi^2|K|$,
whereas the decay into bosons, as was shown in \cite{Enqvist:1998en},
is enhanced by a factor $f_s \simeq 170/|K|^{1/2}$.\footnote{Note that
  the actual decay rate for the Q-giant scenarios is much smaller
  initially until the gas between the Q-giants decay.} From the above,
one obtains an upper bound on the decay width of the
Q-giants\footnote{The reader should note that finite temperature
  decay, as was computed in \cite{Laine:1998rg} is not relevant in our
  scenario since the Q-giants are very weakly coupled to the light
  particles.}
\begin{equation}
  \label{eq:30}
  \Gamma_{Q} = \frac{1}{E_Q}\frac{dE_Q}{dt} \leq \Gamma_Q^{\rm max} \simeq
  \frac{f_s}{24\pi}\, \frac{\sqrt{|K|}m_\phi}{Q}.
\end{equation}

Let us now evaluate the temperature outside during and at the end of
the reheating process.  To do so, we solve the Boltzmann equations, 
\begin{eqnarray}
  \label{eq:49}
  \frac{d \rho_\gamma}{dt} + 4 H \rho_\gamma &=& \int_{Q_{\rm min}}^{Q_{\rm
      max}}dQ f(t,Q)\Gamma_Q^{\rm max} M(Q),
  \\
  \label{eq:50}
  \frac{\partial f(t,Q)}{\partial t} + 3H f(t,Q) &=& Q \Gamma_Q^{\rm max}
  \frac{\partial f(t,Q)}{\partial Q}.
\end{eqnarray}
where $f(t,Q)= \int d^3v f(t,Q,v)$ is the time-dependent density distribution function for
a Q-ball with charge Q, $\rho_\gamma$ is the energy density stored in
radiation and $M(Q) \simeq Qm_\phi$ is the mass of the Q-ball (see
eq. (\ref{eq:25})).  The last term in eq. (\ref{eq:50}) is a transport
term in Q-space, which takes into account the fact that as the Q-giant
decays, it loses its charge gradually. The solution for (\ref{eq:50}) is given by,
\begin{eqnarray}
  \label{eq:51}
  f(t,Q) =  \left(\frac{a}{a_0}\right)^{-3} f\left(t_0,
  Q + \hat\Gamma_Q^{\rm max}(t-t_0)\right)
\end{eqnarray}
where $\hat\Gamma_Q^{\rm max} = Q\Gamma_Q^{\rm max} \simeq m_\phi$ is
independent of $Q$.  Thus, $f(t,Q)$ is simply the initial Gaussian
distribution with a shift of its central value, $\langle Q\rangle
\rightarrow \langle Q\rangle - \hat\Gamma_Q^{\rm max}(t-t_0)$.  This
solution is easy to understand.  For small $t$, $t-t_0 \ll \langle
Q\rangle /\hat\Gamma_Q^{\rm max} = (\Gamma_{\langle Q\rangle}^{\rm max})^{-1}$, the
distribution is approximately constant and the Q-giants redshift like
matter.  At $H^{-1} = t-t_0 = (\Gamma_{\langle Q\rangle}^{\rm max})^{-1}$ the center of
the distribution becomes negative where the solution is no longer
valid.  At this point, the Q-giants decay instantaneously transferring
the energy to radiation.

Next, we solve for the energy density in radiation.  The solution to
(\ref{eq:49}) is
\begin{equation}
  \label{eq:53}
  \rho_\gamma = \left(\frac{a(t)}{a_0}\right)^{-4}\int_{t_i}^{t}dt^\prime
  \left(\frac{a(t^\prime)}{a_0}\right)^4  \int_{Q_{\rm min}}^{Q_{\rm
      max}}dQ f(t^\prime,Q)\Gamma_Q^{\rm max} M(Q).
\end{equation}
Using the above approximation of constant $f(t,Q)$, we get for $t-t_0 \ll (\Gamma_{\langle Q\rangle}^{\rm max})^{-1}$,
\begin{eqnarray}
  \label{eq:55}
  \rho_\gamma \simeq \frac{3(1-\epsilon) m_\phi^5 |K|^{3/2}}{5H_0}\left(\frac{a}{a_0}\right)^{-4}\left[\left(\frac{a}{a_0}\right)^{5/2}-1\right]
\end{eqnarray}
where we took $Q_{\rm min} = 0$ and $Q_{\rm max} = \infty$, and $H_0$
is given in (\ref{eq:1}).  Thus at the transient time, radiation
reheats to a maximal temperature,
\begin{equation}
  \label{eq:37}
  T_{\rm max} \simeq  10^3 \left(\frac{m_\phi}{10^3\textrm{ GeV}}\right)
  \left(\frac{\alpha}{30}\right)^{1/4} \left(\frac{|K|}{10^{-2}}\right)^{1/8}
  \left(\frac{g_*}{200}\right)^{-1/4}\textrm{ GeV},
\end{equation}
and redshifts like $a^{-3/2}$ until $t \simeq \left(\Gamma^{\rm max}_{\langle Q\rangle}\right)^{-1}$ when the Q-giants decay
completely, reheating the universe to
\begin{equation}
  \label{eq:52}
  T_{\rm reheat} = \left(\frac{m_\phi\Mp}{g_*^{1/2}Q}\right)^{1/2} =
  10^{-4} \left(\frac{m_\phi}{10^3\textrm{ GeV}}\right)^{1/2}
  \left(\frac{Q}{10^{28}}\right)^{-1/2}
  \left(\frac{g_*}{200}\right)^{-1/4}\textrm{ GeV}.
\end{equation}
Once the Q-giants decayed, radiation redshifts like $a^{-4}$ as
usual.

As was shown in \ref{sec:stability-gas-phi}, the intermediate reheat
temperature from the decay of the gas is some nine orders of magnitude
above the reheat temperature of the Q-giants.  This, as we advocated
above, is sufficient to allow the Q-giants to regain domination over
the energy density, lost due to collisions.  Furthermore, given the
bound on dissociation, eq. (\ref{eq:36}), we thus find that during the
reheating process, Q-giants are not dissociated and the final reheating
temperature is very low due to their large charge.
Hence, the Q-giants dominate the energy density
too long, leading to a too low reheating temperature for a successful
conventional big bang nucleosynthesis required in the standard
cosmological scenario.  We are therefore led to conclude that negative
pressure flat directions cannot play a role in modular inflation
models.

\section{Positive Pressure Flat Directions and Primordial Black Holes}
\label{sec:posit-press-black}

We now turn to flat directions which, due to RG running, generate
positive pressure, namely, $K > 0$.  Such flat directions do not
fragment to form Q-balls, but rather generate a finite Jeans scale
which prevents the formation of primordial black hole after
inflation  \cite{Carr:1994ar}.  Indeed since typically the
modulus that drives inflation is weakly coupled, there is a long stage
of matter domination after inflation which can generate large density
perturbations and lead to the creation of BHs  \cite{Berkooz:2004yc}.
It is interesting that within the MSSM, the only direction which is
not charged under B-L and which has positive pressure is the $H_uH_d$
direction.

Consider again the potential (\ref{eq:18}), only this time with
positive $K$.  As is shown in Appendix \ref{sec:viri-theor-equat}, the
equation of state is approximately given by,
\begin{eqnarray}
  \label{eq:2}
  P = \frac{K}{K+2}\rho \simeq \frac{K}{2}\rho
\end{eqnarray}
and the Jeans scale is therefore
\begin{equation}
  \label{eq:88}
  \lambda_J = \frac{K}{2} H^{-1}.
\end{equation}
At scales smaller than $\lambda_J$ pressure prevents growth of
perturbations and such perturbations oscillate.  Thus whether or not
BHs can form, depends on the value of $K$ at the time density
perturbations becomes non-linear.  $K$ is given by
\begin{equation}
  \label{eq:3}
  K = \left.\frac{1}{m_\phi^2}\frac{\partial m_\phi^2}{\partial
  t}\right|_{t=\log (|\phi|^2/M^2)}
\end{equation}
with $t = \log (\mu^2/M^2)$ and can therefore be extracted from the RGE's of a given model.  In
what follows, we assume the MSSM up to the GUT scale.  Our strategy is
to find the Jean's scale at the time where $\delta\rho/\rho$ is of
order one and compare it with the scale of the fluctuations.  If the
Jean's scale is larger we are assured that no Q-balls or BHs can form.

Assuming $\delta\rho/\rho \simeq 5\times 10^{-4}$ for scales of order
the horizon at the end of inflation, and using the fact that during
matter domination, perturbations grow like $a$ while $\phi$ redshift
like $a^{-3/2}$, we find
\begin{equation}
  \label{eq:4}
  \mu = \left.|\phi\right|_{\delta\rho/\rho = 1} =
  \frac{\Mp}{(5\times 10^{-4})^{-3/2}} \simeq 10^{13}{\rm GeV}.
\end{equation}
On the other hand, a density fluctuation of  size $\lambda_{\rm inf}
\sim H^{-1}_{\rm inf}$ at the end of inflation becomes when
$\delta\rho/\rho =1$,
\begin{equation}
  \label{eq:6}
  \lambda|_{\delta\rho/\rho=1} = a^{-1/2}H^{-1} \simeq
  2.2\times10^{-2}H^{-1}.
\end{equation}
Thus to a-priori eliminate the possibility of BH formation we must
look for a flat direction of the MSSM with $K/2 \gsim 2.2\times
10^{-2}$.

The RGE's for the MSSM are given in  \cite{Martin:1993zk}. For the
particular case of $H_uH_d$ flat direction, one finds\footnote{We
  thank Guy Englehard for enabling us to use his Mathematica code for
  the RGE's.} $K/2 \simeq 2.3\times 10^{-2}$.  We can therefore
conclude that no primordial BHs can form if the $H_uH_d$ direction
plays the role of the modulus in inflation.  Moreover, it is the only
possible direction in the MSSM.  Before structure can form, the
modulus must decay and reheat the MSSM.  After that, structure
formation can follow as usual, in the radiation dominated universe.

Note that we have imposed a very strong restriction, which is
fortunately fulfilled in the MSSM. It could well be that
$\delta\rho/\rho\simeq 1$ is permissible, in which the universe
undergoes a period of structure formation by $\phi$ particles, similar to
cold dark gas. It will be interesting to explore this possibility
further.

\section{Discussion and Caveats}
\label{sec:discussioncandcaveats}

In this section, we briefly detail what would go wrong with
cosmological scenarios in which the negative pressure flat directions
survive and argue that thermal inflation
\cite{Lyth:1995ka,Stewart:1996ai,Asaka:1999xd} \footnote{We thank Ewan
Stewart for drawing our attention to this issue.} may change the
conclusions. However, we present reasons for the belief that invoking
thermal inflation in the present context requires fine tuning.

With Q-giants surviving a long time, big bang nucleosynthesis would be
severely disrupted since the historical sequence of radiation
domination (RD) and matter domination (MD) periods will be modified
from the standard scenario of
\begin{equation}
\mbox{inflation}+ \mbox{MD} \rightarrow \mbox{RD(BBN)} \rightarrow \mbox{MD}
\rightarrow \mbox{today}
\end{equation}
to 
\begin{equation}
\mbox{inflation} + \mbox{MD} \rightarrow \mbox{RD} \rightarrow \mbox{MD(BBN)}
\rightarrow \mbox{RD} \rightarrow \mbox{MD} \rightarrow \mbox{today}
\label{eq:modified1}
\end{equation}
or
\begin{equation}
\mbox{inflation} +\mbox{MD} \rightarrow
\mbox{RD}\rightarrow \mbox{thermal inflation} + \mbox{MD}\rightarrow \mbox{RD
(BBN)} \rightarrow \mbox{MD} \rightarrow \mbox{RD} \rightarrow
\mbox{MD} \rightarrow \mbox{today}\label{eq:modified2}.
\end{equation}
Eq.~(\ref{eq:modified1}) depicts the situation in which there is no
thermal inflation, while Eq.~(\ref{eq:modified2}) depicts the situation
in which there is thermal inflation to dilute possibly too large
baryon number (note that we have included any possible additional
period of matter domination between thermal inflation and radiation
domination following from flaton decay).  The label ``BBN'' marks
the period during which nucleosynthesis occurs.  A successful
nucleosynthesis is impossible by the time of last RD periods labeled
in Eqs.~(\ref{eq:modified1}) and (\ref{eq:modified2}) since the
maximal reheating temperature of $0.1$ MeV during those time periods
is smaller than $\sim 10$ MeV required for chemical equilibrium
initial conditions or $\sim 1$ MeV when neutrons drop out of
equilibrium \cite{Kolb:vq}.

Clearly, without thermal inflation BBN occurs during matter domination
and thus the existence of Q-giants is excluded.  One can also show
that the baryon-to-entropy ratio during BBN in that case does not
allow for a successful production of the light elements.  We hence
consider the scenario depicted by Eq.~(\ref{eq:modified2}) in which
one invokes a short period of thermal
inflation\cite{Lyth:1995ka,Stewart:1996ai,Asaka:1999xd}. Assume that
we have the flaton, $\varphi$ (not to be confused with $\phi$), at
some VEV, $M$, initially at its true vacuum.  Once the gas decays,
reheating the outer region of the Q-giants to a maximal temperature of
$T_{\rm begin} = T_{\rm inter} = 10^5$ GeV, thermal corrections drive $\varphi$ to the
origin, and thermal inflation begins at roughly that temperature.  It
ends when $T_{\rm end} = m_\varphi \simeq 10^3$ GeV and thus there is at
most 5 e-folds of thermal inflation, which can reduce the energy
density in the Q-giants by $10^{-6}$.

The energy density during thermal inflation is of order
$V_0 = m_\varphi^2M^2$.  Since the flaton stores no energy when it sits in the
true minimum, the total energy density once the gas decays is of order
$g_*T_{\rm inter}^4$ and therefore $V_0$ must be smaller, giving,
\begin{equation}
  \label{eq:Mupperbd}
  M \lesssim 10^8 \left(\frac{m_\varphi}{10^3\textrm{GeV}}\right)^{-1}\left(\frac{g_*}{200}\right)^{1/2} {\rm GeV}.
\end{equation}  
The above can be understood in a different way, by noting that at $M$,
$\varphi$ is coupled to radiation through massive particles with mass
$M$.  Thus corrections to its mass are expected to be of order
$T^4/M^2$ and therefore, for these corrections to drive it to zero,
(\ref{eq:Mupperbd}) must hold.

Within the context of gravity mediated supersymmetry breaking one can
argue that such a requirement from thermal inflation is difficult to
achieve\cite{Lyth:1995ka}.  Indeed, the flaton is expected to have 
some $\mathbb{Z}_n$ symmetry ($n>2$) with a superpotential,
\begin{equation}
  \label{eq:100}
  W = \frac{\varphi^n}{\Mp^{n-3}} + ...
\end{equation}
Taking into account soft breaking mass $-\m^2\varphi^2$ and A-terms
$\m W$ one finds the minimum of the potential to be at
\begin{equation}
  \label{eq:101}
  M = \langle\varphi\rangle \simeq \Mp\left(\frac{\m}{\Mp}\right)^{1/(n-2)}.
\end{equation}
Hence the only value of $M$ which is compatible with
(\ref{eq:Mupperbd}) is $M = \m \simeq 10^3$ GeV for $n=3$.  However,
in such a case, thermal inflation begins only at $T_{\rm begin} = 10^3
= T_{\rm end}$ and so there is no time for thermal inflation to dilute
the Q-giants.  {\it Thus invoking a period of thermal inflation in
  this scenario requires fine tuning}.  Nevertheless, for
completeness, let us suppose that one is able to produce such an epoch
and see what are the requirements for thermal inflation to
sufficiently dilute the Q-giants.

At the end of thermal inflation, assuming perturbative decay of the
flaton, the energy density in radiation is of order $\rho_\gamma^{\rm flat-decay}(T_{\rm flat}) = \gamma m_\varphi^2 M^2$
where $\gamma$ is a factor coming from the volume dilution during the
coherent oscillation of the flaton and $10^{-2} \mbox{ GeV}<T_{\rm
  flat} < 10^3 \mbox{ GeV}$ is the
flaton's reheating temperature (note that the lower bound
on the flaton decay temperature comes from the requirement of
successful BBN).  On the other hand, the energy density in Q-giants is
(including collisions - see section \ref{sec:q-ballq-ball}),
\begin{equation}
\rho_Q(T_{\rm flat}) \simeq 10^{-8}
\gamma g_*(T_{\rm begin}) T_{\rm begin}^4.
\label{eq:qgiantenergyafterthermal}
\end{equation}  
After the flaton decays, in principle, there could occur a baryogenesis
scenario and then BBN can commence afterwards.  Now, for BBN not to be
disrupted, we must have radiation domination during that time.  This
implies,
\begin{equation}
  \label{eq:96}
  M \gsim 10^{6.5}
  \left(\frac{m_\varphi}{10^3\textrm{GeV}}\right)^{-1}\left(\frac{T_{\rm
  flat}}{10^3\textrm{GeV}}\right)^{1/2}\left(\frac{g_*}{200}\right)^{1/2} {\rm GeV}.
\end{equation}
The lower bound decreases for smaller $T_{\rm flat}$
because the flaton decay radiation energy density dilutes faster with
the scale factor than when the flaton energy density is stored in the
form of coherent oscillations.  In the limit that the flaton never
decays but remains in coherent oscillations, since the Q-giants energy
density dilutes the same way as that of the flaton, they
never come to dominate the energy density.
Eqs.~(\ref{eq:Mupperbd}) and (\ref{eq:96}) mean that in
principle, thermal inflation can allow those flat directions not
carrying B to survive.

If the Q-giants carry B charge\footnote{Note that phenomenologically, it is
more appropriate to classify the scenarios according to baryon number
since the observational constraints on lepton number is very weak
\cite{Dolgov:2002wy}.} the situation is more constrained.  In
order not to produce too much baryon number, the Q-giants must never
dominate the energy density.  In such a case it is simple to compute
the baryon asymmetry at the time of Q-giant decay, $T_{\rm
  reheat}\simeq 10^{-4}$.  Indeed, at $T_{\rm reheat}$,
\begin{equation}
  \label{eq:97}
  \rho_Q(T_{\rm reheat}) \simeq \rho_Q(T_{\rm flat})\left(\frac{T_{\rm
      reheat}}{T_{\rm flat}}\right)^3, 
\end{equation}
while,
\begin{equation}
  \label{eq:98}
  \rho_\gamma(T_{\rm flat}) \simeq \gamma m_\varphi^2 M^2 \left(\frac{T_{\rm
      reheat}}{T_{\rm flat}}\right)^4.
\end{equation}
Hence using $n_B/s \simeq (\rho_Q/m_\phi)/\rho_\gamma^{3/4}$ and by demanding that the baryon-to-entropy
ratio so produced is smaller than the measured $10^{-10}$, one finds
\begin{equation}
  \label{eq:99}
   M \gsim 10^{9}
  \left(\frac{m_\varphi}{10^3\textrm{GeV}}\right)^{-3/2}\left(\frac{T_{\rm
  flat}}{10^3\textrm{GeV}}\right)^{1/2}\left(\frac{g_*}{200}\right)^{1/2} {\rm GeV}.
\end{equation}
Thus thermal inflation can, in principle, dilute the baryon number
produced by Q-giants decay, if its reheat temperature is between
$10^{-2}{\textrm{ GeV}} \leq T_{\rm flat} \leq 10{\textrm{ GeV}}$.

Finally, consider the flat direction with positive pressure such that
no Q-giants form.  In that case, thermal inflation can acceptably
dilute the baryon number density if
\begin{equation}
M \gsim 10^{10} \left(\frac{T_{\rm RH}}{10^5\textrm{GeV}}\right)^{1/2}\left(\frac{T_{\rm
  flat}}{10^3\textrm{GeV}}\right)^{1/2}\left(\frac{g_{*S}(T_{\rm RH})}{200}\right)^{1/2} {\rm GeV}.
\end{equation}
where $T_{\rm RH}\leq 10^5 $GeV is the reheating temperature after
inflation ends\footnote{Note that the finite density effect applies
independently of whether there is negative pressure or not.  Note
further that the upper bound of $10^5$ GeV on the temperature of the
universe is quite robust in such models where the modulus controls the
energy density.} and just as before, $T_{\rm flat}$ corresponds to the
temperature at which the flaton decays.  For sufficiently small
$T_{\rm RH}$ and $T_{\rm flat}$ one can dilute the produced baryon
number to an acceptable value.  Hence, the flat direction $H_u L$ for
which $K>0$ can survive with sufficient engineering of thermal
inflation.  (Note that if the reheating temperature is sufficiently
low, sphalerons may be sufficiently suppressed to forbid the
conversion of the lepton number into baryon number.  In that case, no
further dilution from thermal inflation may be necessary.)


To conclude this section, we point out that while thermal inflation
can, for a narrow window of the reheating temperature, dilute the
Q-giants sufficiently thus evading the constraints considered in the
previous sections, such a theory requires fine tuning and so seems to
be improbable.  The same conclusion holds for positive pressure flat
directions.  Finally, even for the flat directions that are
generically ``excluded,'' extremely unlikely fine tuning of initial
conditions (such that no charge is produced) may allow them to be
viable.

\section{Conclusions}
\label{sec:conclusions}

It has long been realized that negative pressure flat directions
fragment into Q-balls which are rather stable and may survive the EW
phase transition.  In this paper, we have considered the evolution of
such flat directions that are assumed to play a role in inflation,
and which therefore have VEVs of order the Planck scale at the end of
inflation and which dominate the energy density at that time.  Such
flat directions are found, for example, in hybrid inflation models.
We have shown that within the framework of gravity mediated
supersymmetry breaking, these flat directions fragment into very large
``Q-giants'' which survive nucleosynthesis and are therefore excluded
cosmologically.

The small decay width of the Q-giants is directly related to its large
charge.  However, thermal effects may break and dissociate the
Q-giants long before they decay.  Indeed, a new effect which was not
considered in previous works is the formation and decay of a gas
outside the Q-giants.  During the fragmentation of a flat direction,
$\phi$, not only Q-giants are formed, but also $\phi$ particles in the
form of a non-thermal gas.  Although, the majority of the charge is
stored inside the Q-giants, there is
an order one fraction of energy density stored in the gas.  While
outside the Q-giants $\langle\phi\rangle$ vanishes, the naive
expectation, that the $\phi$-gas decays very quickly and reheats
the universe to high temperature which would dissociate the Q-giants,
is wrong.  The reason is that $\p$ generates large masses to the
particles to which it couples, and therefore its decay is mediated
through these particles with masses of order $\p$.  One finds the
intermediate temperature outside the Q-giants, once the gas decays to
be,
\begin{equation}
  \label{eq:89}
  T_{\rm inter} \simeq 10^5 \textrm{ GeV},
\end{equation}
which is much less than the dissociation temperature, $T_{\rm diss}
\simeq 10^8$ GeV.  Thus the temperature outside the Q-giants is never
high enough to dissociate them.

Taking collisions into account, we have also shown that by the time
the gas decays it dominates the energy density and therefore there is
a short period of radiation domination, before the Q-giants take
control over the energy density until they decay and reheat the
universe to the temperature 
\begin{equation}
  \label{eq:91}
  T_{\rm reheat} \lesssim 10^{-4} \textrm{ GeV}
\end{equation}
which is generically too low of a reheating temperature to be
compatible with cosmology. 

Hence, without further entropy-releasing mechanisms, negative pressure
flat direction scenarios are not viable cosmological scenarios.  As an
example of entropy-releasing mechanism, we have considered thermal
inflation and found that such a scenario requires fine tuning.
Furthermore, such period of inflation must be properly engineered,
with only a narrow window for the reheating temperature that allows
these flat directions to become viable again.

The only flat directions that do not suffer from too low reheating
temperature without thermal inflation are positive pressure flat
directions.  Hence, within the MSSM, the only possible directions are
$H_uH_d$ and $H_uL$.  Since the latter is charged under B-L and if the
reheating temperature is large enough to produce baryon number, it
requires a dilution mechanism like thermal inflation (which again
requires fine tuning). Finally, we point out that these directions
prevent the formation of primordial BHs due to a generation of a
finite Jeans scale.

\section{Acknowledgement}

We would like to thank M.~Dine, N.~Itzhaki, A.~Kusenko, D.~Kutasov,
and L.~Wang for useful discussions.  We also thank Y.~Nir and
L.~Everett for comments on the manuscript.  The work of MB is
supported in part by the Israel Science Foundation, by the
Braun-Roger-Siegl foundation, by the European network
HPRN-CT-2000-00122, by a grant from the G.I.F.  (the German-Israeli
Foundation for Scientific Research and Development), by the Minerva
Foundation, by the Einstein Center for Theoretical Physics and the by
Blumenstein foundation.

\appendix

\section{The Virial Theorem and Equation of State}
\label{sec:viri-theor-equat}

We present here a simple derivation to the pressure generated by a
scalar condensate oscillating in the potential (\ref{eq:18}) which can also
be approximated by
\begin{equation}
  \label{eq:79}
  U = m_\phi^2\frac{|\phi|^{2K+2}}{M^{2K}}.
\end{equation}
This result was first derived in  \cite{Turner:1983he}.

To do so we consider an alternative version of the virial theorem.  
Consider the quantity,
\begin{equation}
  \label{eq:80}
  \langle\dot F\rangle = \langle\frac{d}{dt}(\phi\dot\phi^\dagger + \dot\phi\phi^\dagger)\rangle
\end{equation}
where the average is taken with respect to time.  Hence, as long as
the motions are bounded, we have
\begin{eqnarray}
  \label{eq:81}
  \langle \dot F\rangle &=&
  \lim_{T\rightarrow\infty}\frac{1}{T}\int_0^T\dot Fdt
  \nonumber \\
  &=& \lim_{T\rightarrow\infty}\frac{1}{T}\left[F(T)-F(0)\right] = 0.
\end{eqnarray}
Thus we can write,
\begin{eqnarray}
  \label{eq:82}
  0 &=& \langle 2|\dot\phi| + \phi\ddot\phi^\dagger +
  \ddot\phi\phi^\dagger\rangle
  \nonumber \\
  &=& \langle 2|\dot\phi| - \phi\partial_\phi U -
  \phi^\dagger\partial_{\phi^\dagger}U\rangle
  \nonumber \\
  &=& \langle 2|\dot\phi| - 2m_\phi^2|\phi|^2(1+K+K\ln\frac{|\phi|^2}{M^2})\rangle
\end{eqnarray}
where we have used the potential (\ref{eq:18}) and ignored the
expansion of the universe.  This is justified assuming the oscillation
period is much smaller than $H^{-1}$.  We therefore find,
\begin{equation}
  \label{eq:83}
 \langle |\dot\phi|\rangle = m_\phi^2\langle|\phi|^2(1+K+K\ln\frac{|\phi|^2}{M^2})\rangle.
\end{equation}

Next, we have,
\begin{eqnarray}
  \label{eq:85}
  \rho &=& \langle|\dot\phi|^2 + U\rangle =
  m_\phi^2\langle|\phi|^2(2+K+2K\ln\frac{|\phi|^2}{M^2})\rangle,
  \\
  P &=& \langle|\dot\phi|^2 - U\rangle =
  m_\phi^2K \langle|\phi|^2\rangle,
\end{eqnarray}
which gives us
\begin{equation}
  \label{eq:86}
  \frac{P}{\rho} =  \frac{K}{2+K+\langle2K|\phi|^2\ln\frac{|\phi|^2}{M^2}\rangle/\langle|\phi|^2\rangle}.
\end{equation}
Similarly, using the potential (\ref{eq:79}) on finds,
\begin{equation}
  \label{eq:87}
  \frac{P}{\rho} =  \frac{K}{2+K}.
\end{equation}


\begin{thebibliography}{9}
  
\bibitem{Sasaki:1995aw}
  M.~Sasaki and E.~D.~Stewart,
  Prog.\ Theor.\ Phys.\  {\bf 95}, 71 (1996)
  [arXiv:astro-ph/9507001].

\bibitem{Chung:2003fi}
  D.~J.~H.~Chung, L.~L.~Everett, G.~L.~Kane, S.~F.~King, J.~Lykken and L.~T.~Wang,
  Phys.\ Rept.\  {\bf 407}, 1 (2005)
  [arXiv:hep-ph/0312378].

\bibitem{Lee:1991ax}
T.~D.~Lee and Y.~Pang,
Phys.\ Rept.\  {\bf 221}, 251 (1992).

\bibitem{Coleman:1985ki}
S.~R.~Coleman,
Nucl.\ Phys.\ B {\bf 262}, 263 (1985)
[Erratum-ibid.\ B {\bf 269}, 744 (1986)].

\bibitem{Lee:1988ag}
  K.~M.~Lee, J.~A.~Stein-Schabes, R.~Watkins and L.~M.~Widrow,
  Phys.\ Rev.\ D {\bf 39}, 1665 (1989).

\bibitem{Kusenko:1997zq}
A.~Kusenko,
Phys.\ Lett.\ B {\bf 405}, 108 (1997)
[arXiv:hep-ph/9704273].

\bibitem{Dvali:1997qv}
  G.~R.~Dvali, A.~Kusenko and M.~E.~Shaposhnikov,
  Phys.\ Lett.\ B {\bf 417}, 99 (1998)
  [arXiv:hep-ph/9707423].

\bibitem{Enqvist:1997si}
K.~Enqvist and J.~McDonald,
Phys.\ Lett.\ B {\bf 425}, 309 (1998)
[arXiv:hep-ph/9711514].

\bibitem{Kusenko:1997si}
  A.~Kusenko and M.~E.~Shaposhnikov,
  Phys.\ Lett.\ B {\bf 418}, 46 (1998)
  [arXiv:hep-ph/9709492].

\bibitem{Turner:1983he}
  M.~S.~Turner,
  Phys.\ Rev.\ D {\bf 28}, 1243 (1983).

\bibitem{Enqvist:1998en}
K.~Enqvist and J.~McDonald,
Nucl.\ Phys.\ B {\bf 538}, 321 (1999)
[arXiv:hep-ph/9803380].

\bibitem{Enqvist:2000gq}
K.~Enqvist, A.~Jokinen and J.~McDonald,
Phys.\ Lett.\ B {\bf 483}, 191 (2000)
[arXiv:hep-ph/0004050].

\bibitem{Kasuya:2000wx}
  S.~Kasuya and M.~Kawasaki,
  Phys.\ Rev.\ D {\bf 62}, 023512 (2000)
  [arXiv:hep-ph/0002285].

\bibitem{Enqvist:2003gh}
  K.~Enqvist and A.~Mazumdar,
  Phys.\ Rept.\  {\bf 380}, 99 (2003)
  [arXiv:hep-ph/0209244].

\bibitem{guthrandall}
L.~Randall, M.~Soljacic and A.~H.~Guth,
Nucl.\ Phys.\ B {\bf 472}, 377 (1996)
[arXiv:hep-ph/9512439];
L.~Randall, M.~Soljacic and A.~H.~Guth,
arXiv:hep-ph/9601296.

\bibitem{Dine:1995uk}
  M.~Dine, L.~Randall and S.~Thomas,
  Phys.\ Rev.\ Lett.\  {\bf 75}, 398 (1995)
  [arXiv:hep-ph/9503303].
 
  
\bibitem{Dine:1995kz}
M.~Dine, L.~Randall and S.~Thomas,
Nucl.\ Phys.\ B {\bf 458}, 291 (1996)
[arXiv:hep-ph/9507453].

\bibitem{Linde:1991km}
A.~D.~Linde,
Phys.\ Lett.\ B {\bf 259}, 38 (1991);
A.~D.~Linde,
Phys.\ Rev.\ D {\bf 49}, 748 (1994)
[arXiv:astro-ph/9307002].

\bibitem{Berkooz:2004yc}
  M.~Berkooz, M.~Dine and T.~Volansky,
  Phys.\ Rev.\ D {\bf 71}, 103502 (2005)
  [arXiv:hep-ph/0409226].

\bibitem{Binetruy:1996xj}
  P.~Binetruy and G.~R.~Dvali,
  Phys.\ Lett.\ B {\bf 388}, 241 (1996)
  [arXiv:hep-ph/9606342].

\bibitem{Halyo:1996pp}
  E.~Halyo,
  Phys.\ Lett.\ B {\bf 387}, 43 (1996)
  [arXiv:hep-ph/9606423].

\bibitem{Stewart:1996ey}
  E.~D.~Stewart,
  Phys.\ Lett.\ B {\bf 391}, 34 (1997)
  [arXiv:hep-ph/9606241];
  E.~D.~Stewart,
  Phys.\ Rev.\ D {\bf 56}, 2019 (1997)
  [arXiv:hep-ph/9703232].

\bibitem{Lyth:1995ka}
  D.~H.~Lyth and E.~D.~Stewart,
  Phys.\ Rev.\ D {\bf 53}, 1784 (1996)
  [arXiv:hep-ph/9510204].

\bibitem{Stewart:1996ai}
  E.~D.~Stewart, M.~Kawasaki and T.~Yanagida,
  Phys.\ Rev.\ D {\bf 54}, 6032 (1996)
  [arXiv:hep-ph/9603324].

  \bibitem{Asaka:1999xd}
  T.~Asaka and M.~Kawasaki,
   Phys.\ Rev.\ D {\bf 60}, 123509 (1999)
   [arXiv:hep-ph/9905467].

\bibitem{Lyth:1996im}
  D.~H.~Lyth,
  Phys.\ Rev.\ Lett.\  {\bf 78}, 1861 (1997)
  [arXiv:hep-ph/9606387].

  


\bibitem{Linde:1983gd}
  A.~D.~Linde,
  Phys.\ Lett.\ B {\bf 129}, 177 (1983).




\bibitem{Copeland:1994vg}
E.~J.~Copeland, A.~R.~Liddle, D.~H.~Lyth, E.~D.~Stewart and D.~Wands,
Phys.\ Rev.\ D {\bf 49}, 6410 (1994)
[arXiv:astro-ph/9401011].
 
\bibitem{Stewart:1994ts}
  E.~D.~Stewart,
   Phys.\ Rev.\ D {\bf 51}, 6847 (1995)
   [arXiv:hep-ph/9405389].

   
\bibitem{Lyth:1997ai}
  D.~H.~Lyth,
  Phys.\ Lett.\ B {\bf 419}, 57 (1998)
  [arXiv:hep-ph/9710347].
  
\bibitem{Gherghetta:1995dv}
T.~Gherghetta, C.~F.~Kolda and S.~P.~Martin,
Nucl.\ Phys.\ B {\bf 468}, 37 (1996)
[arXiv:hep-ph/9510370].

\bibitem{Nilles:1982dy}
  H.~P.~Nilles, M.~Srednicki and D.~Wyler,
  Phys.\ Lett.\ B {\bf 120}, 346 (1983);
  J.~P.~Derendinger and C.~A.~Savoy,
  Nucl.\ Phys.\ B {\bf 237}, 307 (1984);
  M.~Drees,
  Int.\ J.\ Mod.\ Phys.\ A {\bf 4}, 3635 (1989);
   J.~R.~Ellis, J.~F.~Gunion, H.~E.~Haber, L.~Roszkowski and F.~Zwirner,
  Phys.\ Rev.\ D {\bf 39}, 844 (1989).


\bibitem{Kolb:2003ke}
  E.~W.~Kolb, A.~Notari and A.~Riotto,
  Phys.\ Rev.\ D {\bf 68}, 123505 (2003)
  [arXiv:hep-ph/0307241].

\bibitem{Yokoyama:2004pf}
  J.~Yokoyama,
  Phys.\ Rev.\ D {\bf 70}, 103511 (2004)
  [arXiv:hep-ph/0406072].


\bibitem{Multamaki:1999an}
  T.~Multamaki and I.~Vilja,
  Nucl.\ Phys.\ B {\bf 574}, 130 (2000)
  [arXiv:hep-ph/9908446].

\bibitem{Kasuya:2001hg}
  S.~Kasuya and M.~Kawasaki,
  Phys.\ Rev.\ D {\bf 64}, 123515 (2001)
  [arXiv:hep-ph/0106119].

\bibitem{Affleck:1984fy}
I.~Affleck and M.~Dine,
Nucl.\ Phys.\ B {\bf 249}, 361 (1985).

\bibitem{Carena:1996wj}
M.~Carena, M.~Quiros and C.~E.~M.~Wagner,
Phys.\ Lett.\ B {\bf 380}, 81 (1996)
[arXiv:hep-ph/9603420].

\bibitem{Farrar:1996cp}
G.~R.~Farrar and M.~Losada,
Phys.\ Lett.\ B {\bf 406}, 60 (1997)
[arXiv:hep-ph/9612346].

\bibitem{deCarlos:1997ru}
B.~de Carlos and J.~R.~Espinosa,
Nucl.\ Phys.\ B {\bf 503}, 24 (1997)
[arXiv:hep-ph/9703212].

\bibitem{Laine:2000rm}
M.~Laine and K.~Rummukainen,
Nucl.\ Phys.\ B {\bf 597}, 23 (2001)
[arXiv:hep-lat/0009025].

\bibitem{Berkooz:2004kx}
M.~Berkooz, Y.~Nir and T.~Volansky,
arXiv:hep-ph/0401012.

\bibitem{Hall:2005aq}
  L.~J.~Hall, H.~Murayama and G.~Perez,
  arXiv:hep-ph/0504248.

\bibitem{Menon:2004wv}
A.~Menon, D.~E.~Morrissey and C.~E.~M.~Wagner,
arXiv:hep-ph/0404184.

\bibitem{Cohen:1986ct}
  A.~G.~Cohen, S.~R.~Coleman, H.~Georgi and A.~Manohar,
  Nucl.\ Phys.\ B {\bf 272}, 301 (1986).


\bibitem{Multamaki:2000qb}
  T.~Multamaki and I.~Vilja,
  Phys.\ Lett.\ B {\bf 482}, 161 (2000)
  [arXiv:hep-ph/0003270].

\bibitem{Laine:1998rg}
  M.~Laine and M.~E.~Shaposhnikov,
  Nucl.\ Phys.\ B {\bf 532}, 376 (1998)
  [arXiv:hep-ph/9804237].

\bibitem{Carr:1994ar}
  B.~J.~Carr, J.~H.~Gilbert and J.~E.~Lidsey,
  Phys.\ Rev.\ D {\bf 50}, 4853 (1994)
  [arXiv:astro-ph/9405027].

\bibitem{Martin:1993zk}
  S.~P.~Martin and M.~T.~Vaughn,
  Phys.\ Rev.\ D {\bf 50}, 2282 (1994)
  [arXiv:hep-ph/9311340].



\bibitem{Kolb:vq}
E.~W.~Kolb and M.~S.~Turner, {\it The Early Universe},
Redwood City, USA: Addison-Wesley (1990) (Frontiers in physics, 69).

\bibitem{Dolgov:2002wy}
  A.~D.~Dolgov,
  Phys.\ Rept.\  {\bf 370}, 333 (2002)
  [arXiv:hep-ph/0202122].
  
\end{thebibliography}
\end{document}